\documentclass[twocolumn,superscriptaddress,showpacs,prb]{revtex4-1}    
\usepackage{graphicx,amsmath,mathrsfs}
\usepackage{amssymb}
\usepackage{amsthm}
\usepackage{bm}
\usepackage{subfigure,xcolor}   

\def\bc{\begin{center}}
\def\ec{\end{center}}

\newcommand{\bs}[1]{\boldsymbol{#1}}

\newcommand{\nn}{\nonumber}

\newcommand{\pd}{{\phantom{\dag}}}
\newcommand{\up}{\uparrow}
\newcommand{\dw}{\downarrow}
\newcommand{\eps}{\varepsilon}

\def\ie{\emph{i.e.},\ }
\def\eg{\emph{e.g.},\ }

\begin{document}
\title{Spin texture of generic helical edge states}
\author{Alexia Rod}
\affiliation{Institute for Theoretical Physics, Technische Universit\"at Dresden, 01062 Dresden, Germany}
\author{Thomas L. Schmidt}
\affiliation{Department of Physics, University of Basel, Klingelbergstrasse 82, 4056 Basel, Switzerland}
\affiliation{Physics and Materials Science Research Unit,
University of Luxembourg, L-1511 Luxembourg}
\author{Stephan Rachel}
\affiliation{Institute for Theoretical Physics, Technische Universit\"at Dresden, 01062 Dresden, Germany}


\begin{abstract}
We study the spin texture of a generic helical liquid, the edge modes of a two-dimensional topological insulator with broken axial spin-symmetry.
By considering honeycomb and square-lattice realizations of topological insulators, we show that in all cases the generic behavior of a momentum-dependent rotation of the spin quantization axis is realized. Here we establish this mechanism also for disk geometries with continuous rotational symmetry.
Finally, we demonstrate that the rotation of spin-quantization axis remains intact for arbitrary geometries, \ie in the absence of any continuous symmetry.
We also calculate the dependence of this rotation on the model and material parameters. Finally we propose a spectroscopy measurement which should directly reveal the rotation of the spin-quantization axis of the helical edge states.
\end{abstract}


\maketitle

\section{Introduction}

Over the past years, two-dimensional (2D) topological insulators (TIs) have been theoretically predicted and experimentally realized in various semiconductor heterostructures.\cite{hasan10,qi11RMP83_1057} They are two-dimensional systems with a band gap in the bulk material, whose electronic band structure can be characterized by a nontrivial topological $\mathbb{Z}_2$ invariant.\cite{kane-05prl146802} The band gap closes at the interfaces between 2D TIs and topologically trivial systems, \eg the vacuum, so metallic one-dimensional (1D) edge channels must exist at these interfaces.

The edge states of 2D TIs have been studied in detail both experimentally and theoretically. Most importantly, these 1D liquids are \emph{helical}: in a given edge channel, electrons with opposite spins propagate in opposite directions.\cite{kane-05prl146802,kane-05prl226801,bernevig-06s1757} This helicity, which can be observed for instance using the spin Hall effect,\cite{brune10} gives rise to a number of fascinating effects such as unconventional non-local conductances in a multi-terminal setup\cite{roth09} and effective $p$-wave pairing upon inducing superconductivity.\cite{fu09b} Another noteworthy feature of the helical edge states is their robustness to perturbations. Since electrons with opposite momenta are Kramers partners, weak perturbations such as disorder neither lead to a localization of the electron wave functions nor to the opening of a gap in the edge state spectrum, as long as the system is time-reversal invariant.\cite{kane-05prl226801,wu06,xu06} This is in stark contrast to most other 1D liquids.

In most works, it has been implicitly assumed that the electron spin is a good quantum number in the unperturbed edge channel: the very notion that opposite ``spins'' travel in opposite directions is based on this purported symmetry. However, it was pointed out early on that the Hamiltonians governing 2D TIs do not necessarily possess spin symmetry. Indeed, the seminal papers by Kane and Mele\cite{kane-05prl146802,kane-05prl226801} already investigated the effect of Rashba spin-orbit coupling in graphene, which breaks this symmetry. In the 2D TI materials currently under experimental investigation, \ie HgTe/CdTe quantum wells\cite{koenig-07s766} and InAs/GaSb heterostructures,\cite{knez11,knez14} bulk and structural inversion asymmetry both violate the axial spin symmetry of the Hamiltonian. Other promising candidate materials such as silicene\,\cite{mahatha14,vogt12} and tin films\,\cite{yong13} also exhibit intrinsic Rashba spin-orbit coupling breaking the axial spin symmetry.
Naturally, the 1D edge electrons emerging from such 2D Hamiltonians do not have a well-defined spin.

Instead, if the axial spin symmetry is broken, then edge eigenstates with a given momentum $k$ are generally linear combinations of spin-up and spin-down eigenstates with respect to a fixed quantization axis. In that sense, the effect of breaking the axial spin symmetry can be regarded as a $k$-dependent rotation of the spin of the momentum eigenstates. For momenta close to the time-reversal invariant Dirac point ($k=0$ or $k=\pi$), the breaking of the axial spin symmetry in a helical edge state can therefore be described using a single parameter $k_0$, which can be interpreted as the momentum scale on the which spin of the edge electrons rotates.

Zero-energy properties of the edge channels, \eg the linear conductance at zero temperature, are not affected by a perturbation of the axial spin symmetry because they rely only on the fact that electrons with opposite momenta are Kramers partners. Finite-energy properties, in contrast, can indeed be affected because the scattering amplitude between right-moving and left-moving states at different momenta, say $\psi_{-}(k)$ and $\psi_{+}(k')$ ($|k| \neq |k'|$), becomes nonzero. Moreover, since the strength of the symmetry-breaking perturbation can vary in space, devices containing point contacts or tunneling between edge states are sensitive to a nontrivial spin-axis rotation.\cite{orth13}

In this paper, we will therefore investigate in detail the spin structure of helical edge states in time-reversal invariant systems with broken axial spin symmetry. Using analytical as well as numerical methods, we will consider in particular 2D TIs with translation invariance based on the Kane--Mele model or the Bernevig--Hughes--Zhang model in the presence of spin-symmetry breaking terms such as Rashba spin-orbit coupling or bulk inversion asymmetry. Moreover, we will investigate systems with rotational invariance based on these models, which can be experimentally realized in TI disks or flakes.

The structure of this article is as follows: In Sec.~\ref{sec:GHL}, we introduce the notion of \emph{generic helical liquids} (GHL) and motivate the quantities which reveal the rotation of the spin-quantization axis, the main effect considered in this paper. In Sec.~\ref{sec:Models}, we introduce the topological insulator models which we are investigating throughout this work. In the three following sections, the spin texture of the helical edge states and, in particular, the rotation of the spin-quantization axis and its dependence on system parameters, is studied for three different set-ups: (i) exact diagonalization for tight-binding models on nanoribbons (\ie translational invariance in one direction), (ii) analytical solutions for continuum models on circular disks (\ie rotational invariance), and (iii) exact diagonalization for tight-binding models on disks (\ie no continuous symmetry).
Eventually, in Sec.~\ref{sec:spec}, a spectroscopic consideration is discussed for real space disks, before we summarize our findings in Sec.~\ref{sec:conclusions}.

\section{Generic helical liquids}\label{sec:GHL}

Counterpropagating helical states on a given edge of a 2D TI are related by time-reversal symmetry. For a noninteracting translation-invariant edge state, the single-particle momentum $k$ is a good quantum number. If we denote right-moving and left-moving eigenstates of the Hamiltonian by the operators $\psi_+(k)$ and $\psi_-(k)$, these are linked by time-reversal,
\begin{align}\label{eq:TR}
    T \psi_-(k) T^{-1} &= \psi_+(-k), \notag \\
    T \psi_+(k) T^{-1} &= -\psi_-(-k).
\end{align}
and the minus sign in the last line follows from the fact that the anti-unitary time-reversal operator $T$, when acting on single fermions, should satisfy $T^2 = -1$. If the Hamiltonian is time-reversal invariant, $T H T^{-1} = H$, this rules out many common scattering and interaction processes. For instance, single-particle potential backscattering would produce a term containing $\psi^\dag_+(k) \psi_-(k') + \text{h.c.}$ which is not time-reversal invariant.

As the relations (\ref{eq:TR}) are identical to those of spinful fermions, it is often permissible to think of the label $\pm$ as a spin label and to replace $\psi_{+,-}(k)$ by $\psi_{\uparrow,\downarrow}(k)$. If one considers for instance elastic scattering, the observable properties of the system are determined by the overlap between states with equal energy, e.g., $\psi_-(-k)$ and $\psi_+(k)$. These states have zero overlap because they are Kramers partners.

However, if $\psi_\pm(k)$ are indeed spin eigenstates, this orthogonality persists even for states with different momenta: the overlap of $\psi_\uparrow(k)$ with $\psi_\downarrow(k')$ vanishes even for $|k| \neq |k'|$ because of the spin degree of freedom. Importantly, the orthogonality of these states does not follow from time-reversal symmetry. There are several scenarios where this distinction between becomes relevant.

For instance, electron-electron or electron-phonon interactions may cause inelastic scattering. In that case, the overlap of states $\psi_+(k)$ and $\psi_-(k')$ at different momenta $|k| \neq |k'|$ enters the physical observables. If these states are not spin eigenstates, their overlap can be nonzero even if the system is time-reversal invariant. An important consequence is a deviation of the conductance of the helical edge state from the conductance quantum at finite temperatures.\cite{schmidt-12prl156402} Moreover, the spin structure of the edge state can in principle be probed by injecting particles with spin polarization along a given axis and measuring the absorption.

Let us define a fixed but arbitrary spin quantization axis, and let $\psi^\dag_\uparrow(k)$ and $\psi^\dag_\downarrow(k)$ create states with momentum $k$ and given spin projection along this axis. Then, the left- and right-moving eigenstates of the Hamiltonian are in general linear combinations,
\begin{align}
    \begin{pmatrix} \psi_\uparrow(k) \\ \psi_\downarrow(k) \end{pmatrix}
    = B_k
    \begin{pmatrix} \psi_+(k) \\ \psi_-(k) \end{pmatrix}
\end{align}
where $B_k$ is a momentum-dependent unitary $2 \times 2$ matrix. We can choose the spin quantization axis in such a way that $B_0 = 1$ at the Dirac point $k=0$. The fact that the pairs $[\psi_{+}(k), \psi_{-}(-k)]$ and $[\psi_{\uparrow}(k), \psi_{\downarrow}(-k)]$ are both related by time-reversal constrains the form of $B_k$. In particular, one finds that $B_k$ must be an even function of $k$: $B_k = B_{-k}$. Being a unitary matrix, this means that up to an irrelevant phase, the low-momentum behavior of $B_k$ is characterized by a single parameter $k_0$,
\begin{align}
    B_k \approx \begin{pmatrix} 1 & -(k/k_0)^2 \\ (k/k_0)^2 & 1 \end{pmatrix}
\end{align}
up to order $k^2$. Physically, $k_0$ can be interpreted as the characteristic momentum scale for the rotation of the spin quantization axis. Its value directly enters the formula for the temperature-dependent correction to the otherwise quantized edge state conductance, as shown in Refs.~\onlinecite{schmidt-12prl156402,kainaris14}. Moreover, it can be measured using point contacts in narrow topological insulators.\cite{orth13} Since $B_k$ cannot be extracted directly from our numerical simulations, we focus instead on $k_0$. In order to extract $k_0$, we consider the scattering between left-movers and right-movers. In the simplest case of a potential scatterer which is short-ranged compared to the Fermi wavelength of the edge state, but long-ranged compared to the penetration depth of the edge state into the bulk, the scattering amplitude is proportional to expression\cite{schmidt-12prl156402}
\begin{equation}\label{RQA-old}
\left[ B_{k_2}^\dag B_{k_1}^\pd \right]^{-+} = \int dx \,\psi_{-,k_2}^\dag(x,y_0) \, \psi_{+,k_1}^\pd(x,y_0)
\end{equation}
where $\psi_{\pm,k}(x,y)$ denotes the wave function of a right (left) moving edge state with momentum $k$, evaluted at position $(x,y)$. Here, $x$ ($y$) is the coordinate transversal to the edge state (along the edge state). It has further been shown\,\cite{schmidt-12prl156402} that Eq.~\eqref{RQA-old}
behaves at long wavelengths as
\begin{equation}
\left[ B_{k_2}^\dag B_{k_1}^\pd \right]^{-+} \approx  k_0^{-2}\left(k_1^2-k_2^2\right) \label{eq:overlap_cylinder}
\end{equation}
which enables us to conveniently extract $k_0$ from numerical simulations. In practice, a single-particle Hamiltonian matrix is diagonalized for each $k$ individually. Within a numerical diagonalization of a matrix, all eigenvectors will be computed with an arbitrary U(1) phase attached to them (for a given $k$, all eigenvectors have the same phase factor, but for different $k$-values, \ie for different diagonalizations, the phase factors will be different in general). In order to overcome this technical issue, we consider in the following the modulus of $[ B_{k_2}^\dag B_{k_1}^\pd ]^{-+}$ which is gauge-invariant and will be the central object considered in this paper. We define the \emph{ rotation of spin-quantization axis} (RSQA),
\begin{equation}
\label{rsqa}
\begin{split}
\mathcal{K}(k_1,k_2) =&~ \left| \int dx \,\psi_{-,k_2}^\dag(x,y_0) \, \psi_{+,k_1}^\pd(x,y_0) \right| \\[10pt]
\approx & ~ k_0^{-2}\left| k_1^2-k_2^2\right|
\end{split}
\end{equation}
with RSQA amplitude $k_0^{-2}$.
When considering disks with circular shape the momentum quantum numbers $k$ will be replaced  by angular momentum quantum numbers $j$ and an RSQA amplitude $j_0^{-2}$. In the case of two-dimensional flakes (\ie disks not possessing a rotationally invariant shape) we will characterize the edge states simply by the energies $E$, leading to an RSQA amplitude $\eps_0^{-2}$.

The physical relevance of the RSQA (\ref{rsqa}) is twofold: first, the expression for $\mathcal{K}(k_1,k_2)$ occurs in matrix elements of \emph{local} electronic Hamiltonians evaluated in the basis of the helical edge states. Physical quantities which can be related to such matrix elements, e.g., the change in electric or heat conductance due to impurity scattering, or the local tunneling probability into the edge state, will depend on $\mathcal{K}$. Moreover, a numerical evaluation of $\mathcal{K}$ allows us to determine $k_0$, which is a quantity characterizing the edge state spectrum near the Dirac point. It can be measured using nonlocal measurements such as momentum-conserving tunneling between nearby edge states.

\section{Topological Insulator Models}\label{sec:Models}
In the following, we briefly introduce the topological insulator (TI) models which are discussed throughout this paper.

The first model is the Bernevig--Hughes--Zhang (BHZ) model\cite{bernevig-06s1757} which was proposed to describe the TI phase of the HgTe/CdTe quantum wells.\cite{koenig-07s766} Here we consider both the continuum model as well as the square lattice tight-binding version. The breaking of the axial spin symmetry (\ie the spin $S^z$ symmetry) is accomplished by adding a term describing the bulk inversion asymmetry (BIA).

The second TI model is the Kane--Mele model\cite{kane-05prl146802,kane-05prl226801} for the honeycomb lattice which was originally proposed to describe quantum spin Hall effect in graphene. The breaking of axial spin symmetry is realized by a Rashba spin-orbit term (called Rashba 1)  which might be induced by an external electrical field. Then  we also briefly discuss the Kane--Mele model applied to the more realistic TI-candidate materials silicene, germanene, and stanene -- the silicon, germanium, and tin analogues of graphene. Due to the buckled lattice structure of these materials, an additional Rashba spin-orbit term is allowed by symmetries (called Rashba 2) which also breaks the axial spin symmetry.

\subsection{Bernevig--Hughes--Zhang model}

The BHZ model~\cite{bernevig-06s1757,qi11RMP83_1057} in the continuum is defined
in the basis $\{|E_1,+\rangle,|H_1,+\rangle,|E_1,-\rangle,|H_1,-\rangle\}$ with the following matrix:
\begin{equation}
\label{eq:BHZ}
\mathcal{H}_{\rm BHZ}=\begin{pmatrix}
h(\mathbf{k}) & 0\\
0 & h^\star(-\mathbf{k})
\end{pmatrix}
\end{equation}
where $+$ and $-$ can be seen as effective spin up and down, respectively. The $2\times 2$ sub-block associated with a single spin is defined as
\begin{equation}
h(\mathbf{k})=\varepsilon(\mathbf{k})\mathbb{I}_{2x2}+d_i(\mathbf{k})\sigma^i,
\end{equation}
where
\begin{subequations}
\begin{eqnarray}
&&\varepsilon(\mathbf{k})=C-D(k_x^2+k_y^2)\ ,\\[5pt]
&&\mathbf{d}(\mathbf{k})=\Big[Ak_x,-Ak_y,M-B(k_x^2+k_y^2)\Big]\ .
\end{eqnarray}
\end{subequations}
$A$, $B$, $C$ and $D$ are system or material parameters, $M$ describes the band gap
and $\sigma_i$ acts on the orbital subspace $(E,H)$. This model can be regularized on a two-orbital square lattice,\cite{bernevig-06s1757}
\begin{eqnarray}
&&\varepsilon(\mathbf{k})=C-2D[2-\cos(k_x)+\cos(k_y)]\ ,\\[5pt]
&&\mathbf{d}(\mathbf{k})=[A\sin(k_x),-A\sin(k_y),M(\mathbf{k})]\ ,\\[5pt]
&&M(\mathbf{k})=M-2B(2-\cos(k_x)+\cos(k_y))\ .
\end{eqnarray}
Note that the lattice spacing is set to unity throughout the paper.
The corresponding Bloch matrix  $\mathcal{H}_{\rm BHZ}$ from $H_{\rm BHZ}=\sum_{\bs{k}} \Psi^\dag(\bs{k}) \mathcal{H}_{\rm BHZ} \Psi(\bs{k})^\pd$  can be written as\,\cite{koenig-08jpsj031007}
\begin{eqnarray}\label{ham:bloch-bhz}
\mathcal{H}_{\rm BHZ}~=~t&&(A\sin(k_x)\sigma_x\otimes s_z-A\sin(k_y)\sigma_y\otimes \mathbb{I}_{2x2}\nonumber\\[5pt]
&&+M(\mathbf{k})\sigma_z \otimes \mathbb{I}_{2x2} + \varepsilon(\mathbf{k}) \mathbb{I}_{4x4})\ ,
\end{eqnarray}
where  $\Psi(\bs{k})=(e_\up(\bs{k}), h_\up(\bs{k}), e_\dw(\bs{k}), h_\dw(\bs{k}))^T$ is a four-component spinor. $s_i$ denotes the Pauli matrices for physical spin and $\sigma_i$ for the orbital degrees of freedom. Already in Ref.\,\onlinecite{bernevig-06s1757} it was mentioned that a term stemming from BIA will be present which gives the following contribution to the Bloch matrix\,\cite{bernevig-06s1757,qi11RMP83_1057},
\begin{equation}\label{eq:BIA}
\mathcal{H}_{\rm BIA}=\begin{pmatrix}
0 & 0 & 0 & -\Delta\\
0 & 0 & \Delta & 0\\
0 & \Delta & 0 & 0\\
 -\Delta & 0 & 0 & 0\\
\end{pmatrix},
\end{equation}
with a material parameter $\Delta$. The BIA term is important here since it couples the two spin channels and leads to breaking of spin $S^z$ symmetry, \ie spin is not conserved anymore. Another source for breaking of the axial spin symmetry would be Rashba spin-orbit coupling,\cite{rothe-12njp065012} which we do not consider here for the BHZ model.
In order to derive the tight-binding hopping Hamiltonian in real space we fourier-transform \eqref{ham:bloch-bhz} and \eqref{eq:BIA} and obtain
\begin{eqnarray*}
&&\quad H_{\rm BHZ}=(C-4D)\sum_{i\sigma}\left( e_{i\sigma}^\dag e_{i\sigma}^\pd + h_{i\sigma}^\dag h_{i\sigma}^\pd \right) \\
+&& (M-4B)\sum_{i\sigma}
\left( e_{i\sigma}^\dag e_{i\sigma}^\pd - h_{i\sigma}^\dag h_{i\sigma}^\pd \right)\\
+&&
\Bigg[ D\sum_{i\sigma}  \left( e_{i+\hat x \sigma}^\dag e^\pd_{i\sigma} +
e_{i+\hat y \sigma}^\dag e^\pd_{i\sigma} + h_{i+\hat x \sigma}^\dag h^\pd_{i\sigma} +
h_{i+\hat y \sigma}^\dag h^\pd_{i\sigma} \right)\Bigg.\\
+&&
B\sum_{i\sigma}  \left( e_{i+\hat x \sigma}^\dag e^\pd_{i\sigma} +
e_{i+\hat y \sigma}^\dag e^\pd_{i\sigma} - h_{i+\hat x \sigma}^\dag h^\pd_{i\sigma} -
h_{i+\hat y \sigma}^\dag h^\pd_{i\sigma} \right)\\
+&&
\frac{A}{2}\sum_i \left( -i e_{i+\hat x\up}^\dag h_{i \up}^\pd +i e_{i-\hat x\up}^\dag h_{i \up}^\pd
+i e_{i+\hat x\dw}^\dag h_{i \dw}^\pd -i e_{i-\hat x\dw}^\dag h_{i \dw}^\pd
\right)\\
+&&
\frac{A}{2} \sum_{i\sigma} \left(
e_{i+\hat y\sigma}^\dag h_{i \sigma}^\pd - e_{i-\hat y\sigma}^\dag h_{i \sigma}^\pd \right)\\
+&&\Bigg.\Delta \sum_i
\left( e_{i \up}^\dag h_{i \dw}^\pd  - h^\dag_{i\up} e^\pd_{i\dw}  \right) + {\rm h.c.}\Bigg]\ .
\end{eqnarray*}
In order to perform the exact diagonalization on nanoribbons, one has to partially fourier-transform the Hamiltonian such that one ends up with $H_{\rm BHZ}(x,k_y)$ or $H_{\rm BHZ}(k_x,y)$, respectively. Of course, these representations can be obtained by properly combining the previous equations.

\subsection{Kane--Mele model}
%
The Kane--Mele (KM) model\,\cite{kane-05prl146802,kane-05prl226801} was originally proposed to describe the quantum spin Hall effect in graphene. Although the spin-orbit coupling in graphene is far too small to be observed experimentally, the KM model serves as toy model to study two-dimensional TIs on the honeycomb lattice. Moreover, the Kane-Mele papers have stimulated the search for other honeycomb-lattice materials with possibly heavier elements in order to realize strong spin-orbit coupling. Examples are silicene, germanene, and stanene as discussed below.
The tight-binding version of the KM model is governed by the Hamiltonian
\begin{eqnarray}\label{eq:Kane_Mele_tb}
\mathcal{H}_{\rm KM}&=&-t\sum\limits_{\langle ij\rangle}c^\dagger_ic_j+i\lambda_{\rm SO}\sum\limits_{\langle\!\langle ij \rangle\!\rangle}\nu_{ij}c^\dagger_is^zc_j\nonumber\\
&&{}+i\lambda_{\rm R, 1}\sum\limits_{\langle ij \rangle}c^\dagger_i(\bs{s}\times \hat{\bs{d}}_{ij})_zc_j^\pd, 
\end{eqnarray}
where
$c_i=(c_{i\uparrow},c_{i\downarrow})$ is a two-component spinor and $t$ is the hopping constant,
 $\lambda_{\rm SO}$ the intrinsic spin-orbit coupling, and $\lambda_{\rm R,1}$ the Rashba spin-orbit coupling. Moreover, $\nu_{ij}=\pm 1$ is a phase factor and the Pauli matrix $s^\alpha$ denotes the physical spin, $\hat{\bs{d}}_{ij}$ is the unit vector between sites $i$ and $j$.
We further use the following nearest-neighbor vectors on the honeycomb lattice,
\begin{eqnarray}\nn
\delta_1=\dfrac{1}{\sqrt{3}}\begin{pmatrix}
-1\\ 0
\end{pmatrix}\,,~ &  \delta_2 = \dfrac{1}{2\sqrt{3}} \begin{pmatrix}
1 \\ \sqrt{3}
\end{pmatrix}\,,~ & \delta_3=\frac{1}{2\sqrt{3}}\begin{pmatrix}
1 \\ -\sqrt{3}
\end{pmatrix}.
\end{eqnarray}
In momentum space, the KM Hamiltonian reads $H_{\rm KM}=\sum_{\bs{k}} \Phi(\bs{k})^\dag \, \mathcal{H}_{\rm KM} \,\Phi(\bs{k})^\pd$ with the Bloch matrix $\mathcal{H}_{\rm KM}$ and the four-component spinor $\Phi(\bs{k})=(a_\up(\bs{k}), b_\up(\bs{k}), a_\dw(\bs{k}), b_\dw(\bs{k}))^T$. The Bloch matrix is given by
 \begin{equation}
 \mathcal{H}_{\rm KM} = \left(\begin{array}{cccc}\gamma &  -g & \cdot & \chi_1 \\[5pt]
 -g^\ast & -\gamma & \chi_2 & \cdot \\[5pt]
 \cdot & \chi_2^\ast & -\gamma &  -g \\[5pt]
 \chi_1^\ast & \cdot & -g^\ast & \gamma \end{array}\right)\ .
 \end{equation}
 Here we used the following abbreviations $g$, $\gamma$, $\chi_1$, and $\chi_2$, which are $\bs{k}$ dependent functions:
%
%
 \begin{eqnarray}
&&g = t\left[1+e^{\sqrt{3}ik_x/2}e^{ik_y/2}+e^{ik_y}\right]\ , \\[10pt]
&&\gamma = 2\lambda_{\rm SO}\! \left[\sin(k_y)-2\sin(k_y/2)\cos(\sqrt{3}k_x/2)\right]\ ,\\[10pt]
&&\chi_1 = \frac{i\lambda_{R,1}}{2}\left[(-\sqrt{3}-i)+e^{ik_y}(\sqrt{3}-i)\right. \\
&& \nonumber \qquad\qquad\qquad \left. +2ie^{\sqrt{3}ik_x/2}e^{ik_y/2}\right] \\[10pt]
%
&&\chi_2 = \frac{i\lambda_{R,1}}{2}\left[(\sqrt{3}+i)+e^{-ik_y}(-\sqrt{3}+i)\right. \\
&& \nonumber \qquad\qquad\qquad \left. -2ie^{-\sqrt{3}ik_x/2}e^{-ik_y/2}\right]
\end{eqnarray}
While the $g$ term describes the semi-metallic behavior of graphene, the $\gamma$ term leads to the topological insulating phase proposed by Kane and Mele. Note that $\gamma$ was first advocated by Haldane\,\cite{haldane88prl2015} in order to realize quantum Hall effect on the honeycomb lattice.
The $\chi$ terms correspond to Rashba spin-orbit coupling and mix the spin channels leading to a broken axial spin-symmetry. Typically the Rashba term is due to an external electrical field or the effect of a substrate.

\subsection{Silicene, Germanene, Stanene}

As mentioned above, the KM model was introduced to describe a TI phase in graphene, a monolayer of graphite. While graphene does not feature such a phase, climbing up in the periodic table might be a promising route as spin-orbit coupling is expected to be much stronger. Therefore people have considered monolayers of silicon, germanium, and tin, called {\it silicene}, {\it germanene}, and {\it stanene}. In contrast to graphene, these materials have a buckled honeycomb structure, \ie the two sublattices of the honeycomb lattice are not coplanar.
This gives rise to an intrinsic Rashba spin-orbit interaction between next-neareset neighbor sites, which is expected to be much stronger than the Rashba spin-orbit interaction coming from an external electric field or the substrate. The corresponding Hamiltonian is  the KM Hamiltonian \eqref{eq:Kane_Mele_tb}  with additional intrinsic Rashba interaction,
\begin{equation}
H_{{\rm R}_2}=-i\frac{2}{3}\lambda_{\rm R,2}\sum\limits_{\langle\!\langle ij \rangle\!\rangle}\mu_{ij} \,c^\dagger_i(\bs{s} \times \hat{\bs{d}}_{ij})_z\,c_j^\pd,
\end{equation}
where $\mu_{ij}=\pm 1$ for the A (B) sublattice and  $ \hat{\bs{d}}_{ij}$ is the unit vector between the second nearest neighbor sites $i$ and $j$. The corresponding contribution to the Bloch matrix is given by
\begin{equation}
\mathcal{H}_{\rm R,2} = \begin{pmatrix}
\cdot & \cdot & -\chi_3 & \cdot\\
\cdot & \cdot & \cdot & \chi_3\\
-\chi_3^\ast & \cdot & \cdot & \cdot\\
\cdot & \chi_3^\ast & \cdot & \cdot
\end{pmatrix}\end{equation}
with
\begin{equation}
\begin{split}
\chi_3 = &\frac{4}{3}\lambda_{R,2}\left[\sin(k_y)+\cos(\sqrt{3}k_x/2)\sin(k_y/2)\right. \\
&\quad \left.+\sqrt{3}i\sin(\sqrt{3}k_x/2)\cos(k_y/2))\right]
\end{split}
\end{equation}

%
%
\section{Rotation of  spin-quantization axis I: tight-binding ribbons}

\subsection{Set-up}

In the following, we consider both the BHZ and the KM tight-binding models on a nanoribbon geometry which possesses open boundary conditions in one and periodic boundary conditions in the other direction. In order to achieve this one needs to partially Fourier-transform, say in $y$ direction while keeping all hoppings along the $x$ direction in real space. The Bloch wave functions will have the following form:
\begin{figure}[b!]
\centering
\includegraphics[scale=0.65]{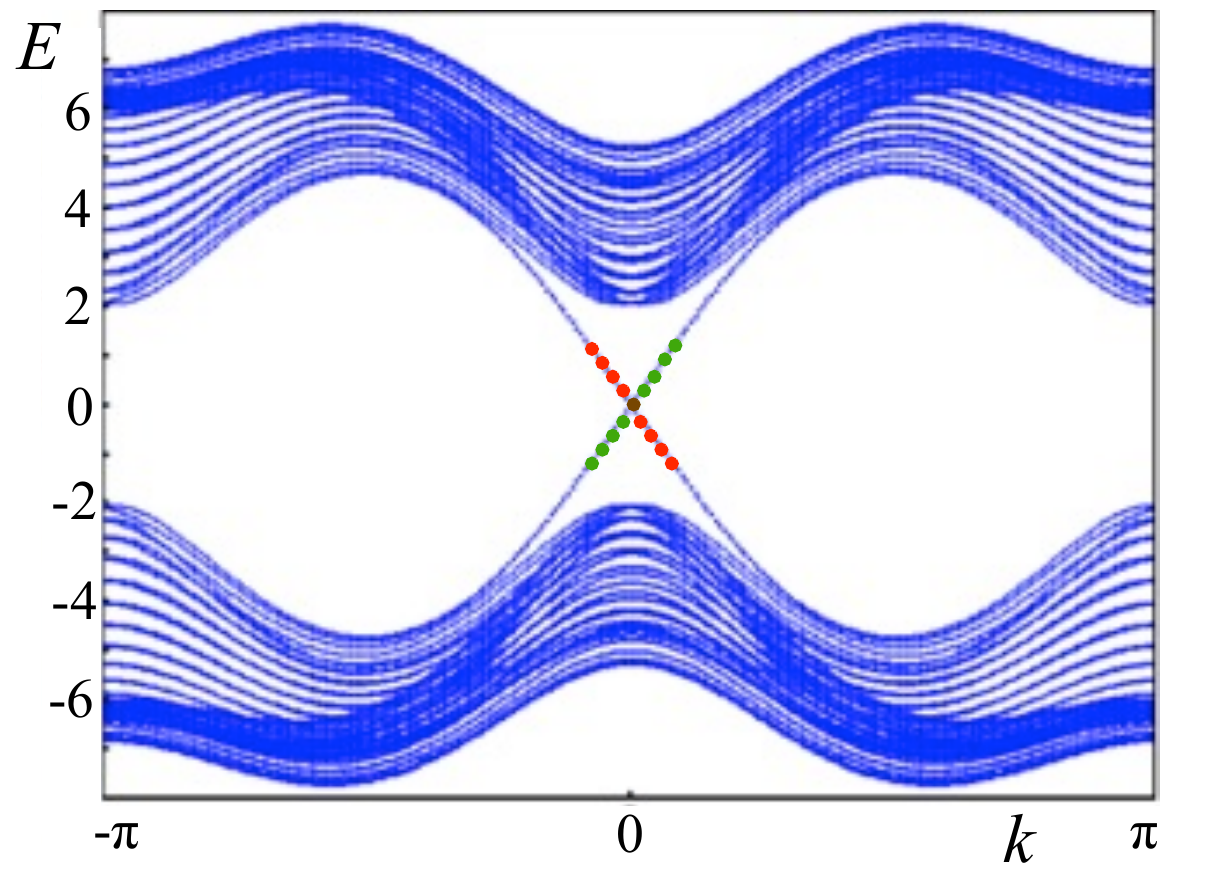}
\caption{(Color online).  Spectrum of the BHZ model on a nanoribbon with $L=24$ unit cells. Parameters: $A=5$, $B=-1$, $M=-2$, $\Delta=0.3$, $C=D=0$.}
\label{fig:bhz-spec}
\end{figure}
\begin{equation}
\psi(x,y) 
=\tilde{\psi}(x,k)e^{iky}. \label{eq:cylinder_wf}
\end{equation}%
%
%
%
%
\begin{figure}[b!]
\centering
\includegraphics[scale=0.65]{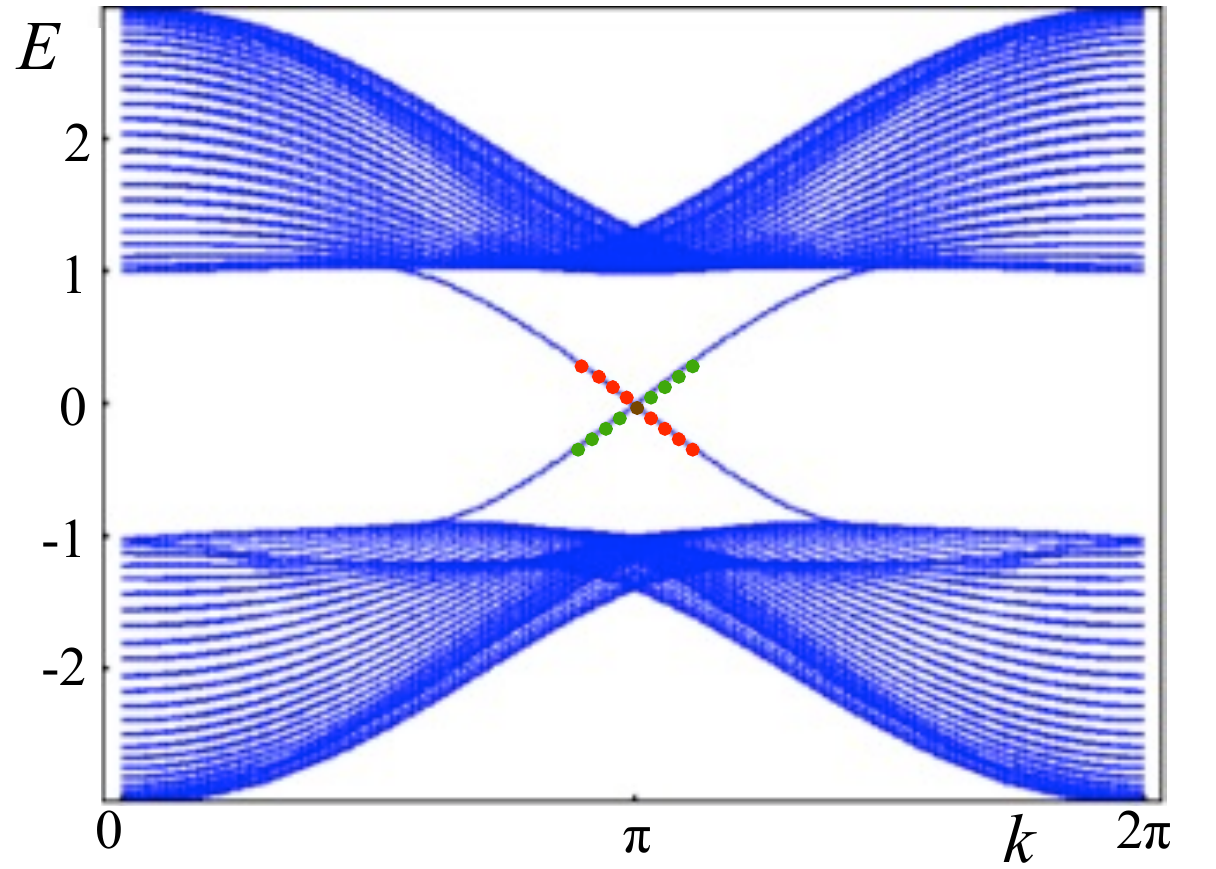}
\caption{(Color online).  Spectrum of the KM model on a nanoribbon with $L=24$ unit cells. Parameters: $t=1$, $\lambda_{\rm SO}=0.2$, $\lambda_{\rm R,1}=0.05$, and  $\lambda_{\rm R,2}=0$.}
\label{fig:kmr-spec}
\end{figure}

\begin{figure*}[t]
\includegraphics[width=0.34\linewidth]{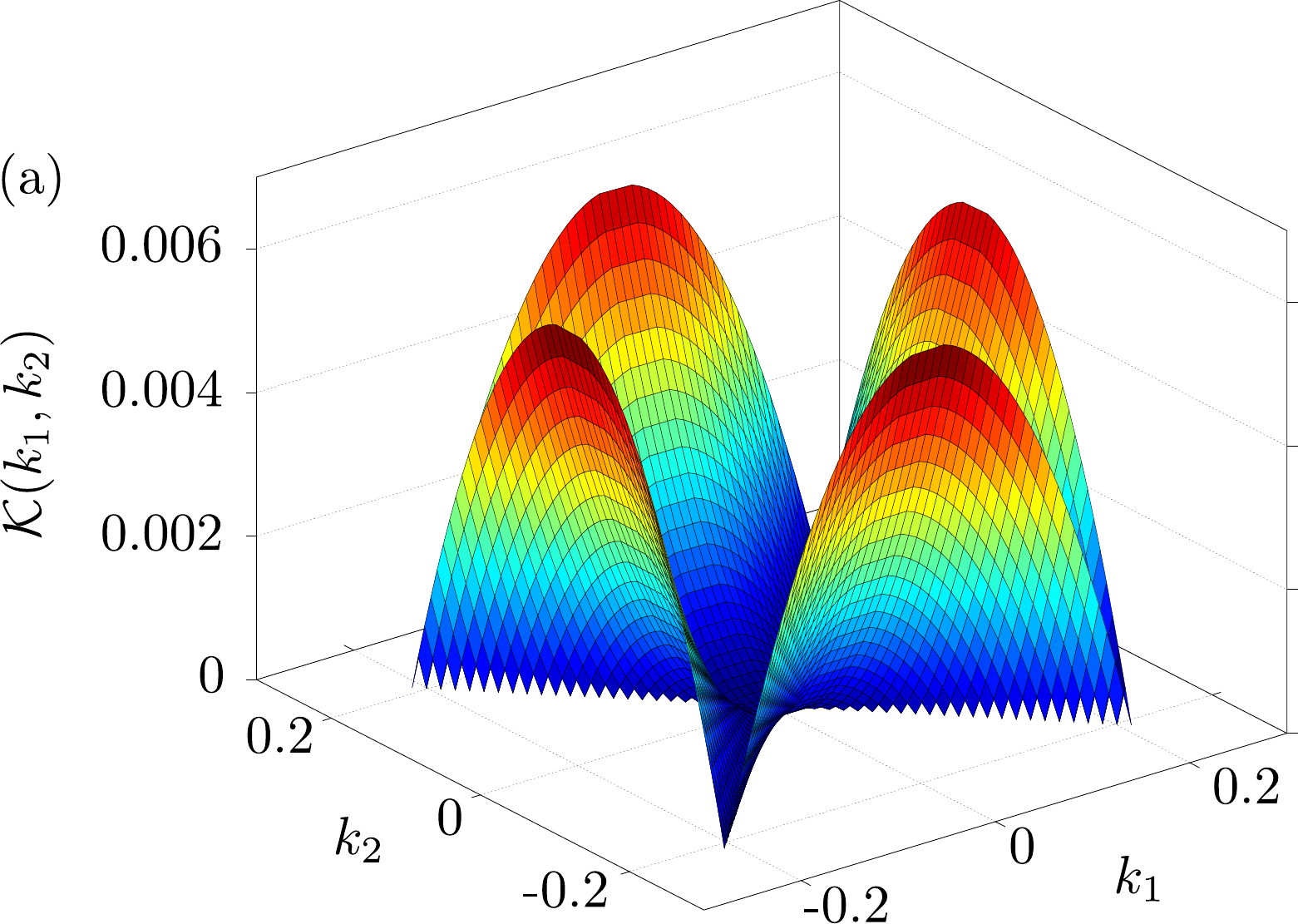}
\hspace{0.01\linewidth}
\includegraphics[width=0.3\linewidth]{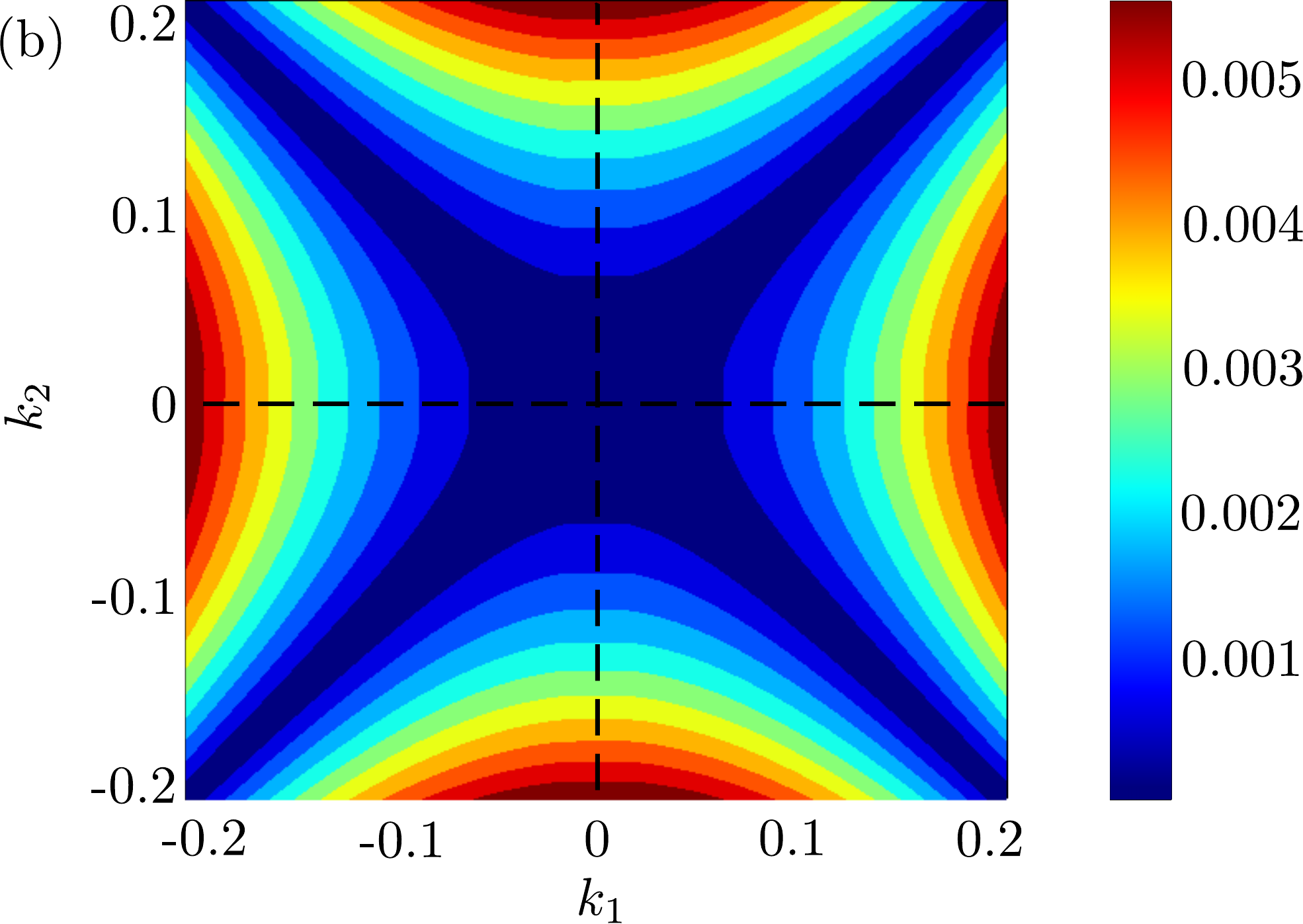}
\hspace{0.01\linewidth}
\includegraphics[width=0.31\linewidth]{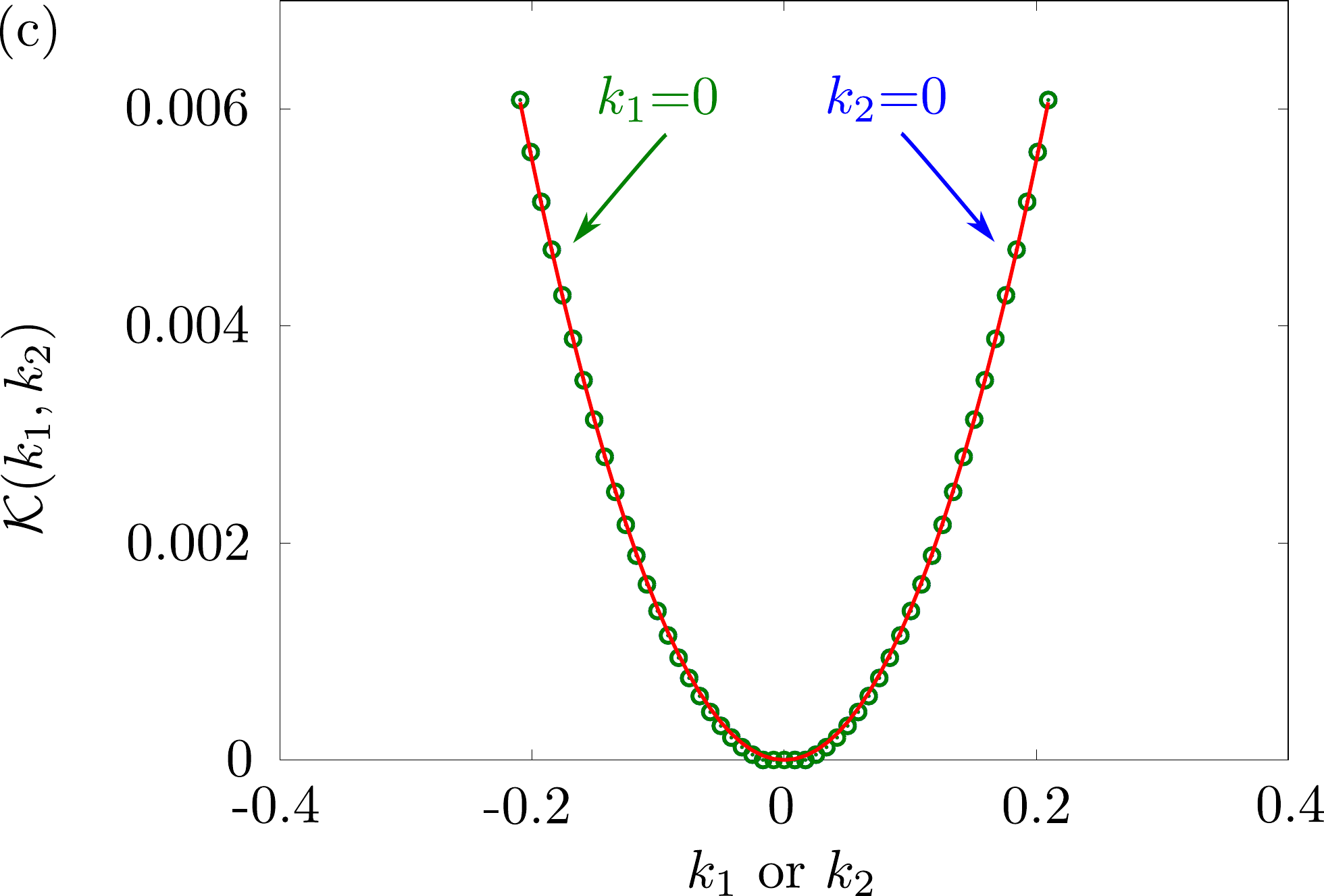}
\includegraphics[width=0.34\linewidth]{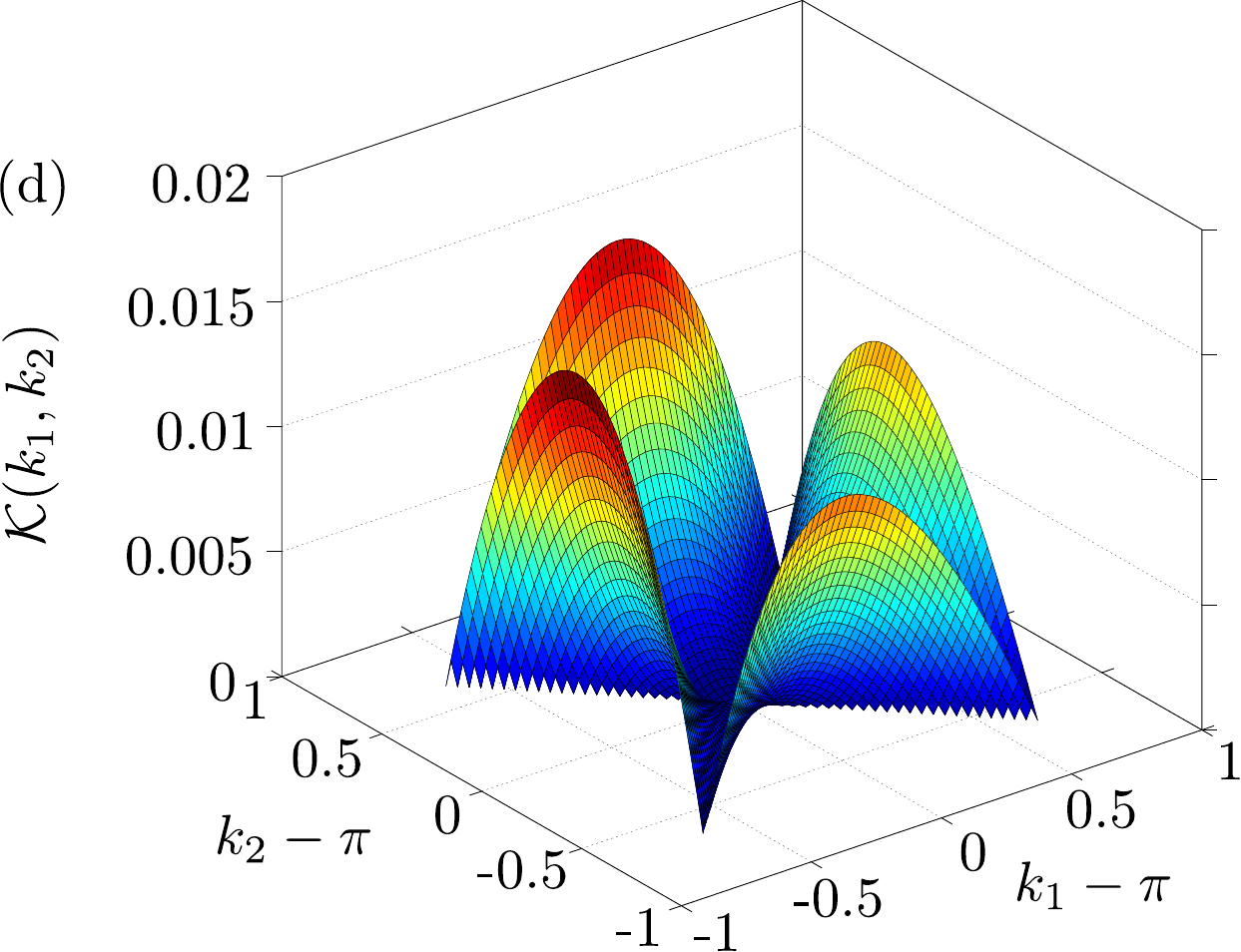}
\hspace{0.01\linewidth}
\includegraphics[width=0.3\linewidth]{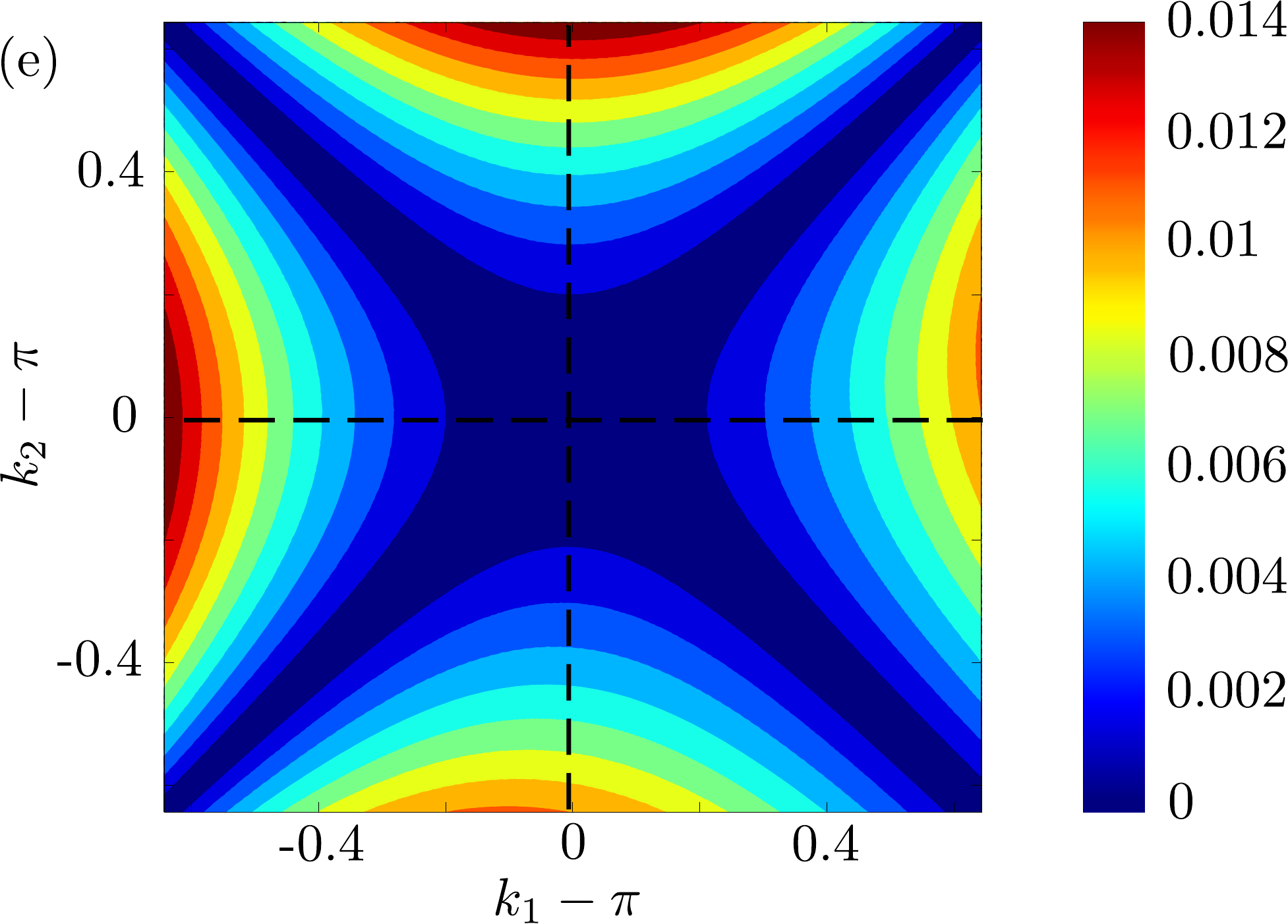}
\hspace{0.01\linewidth}
\includegraphics[width=0.3\linewidth]{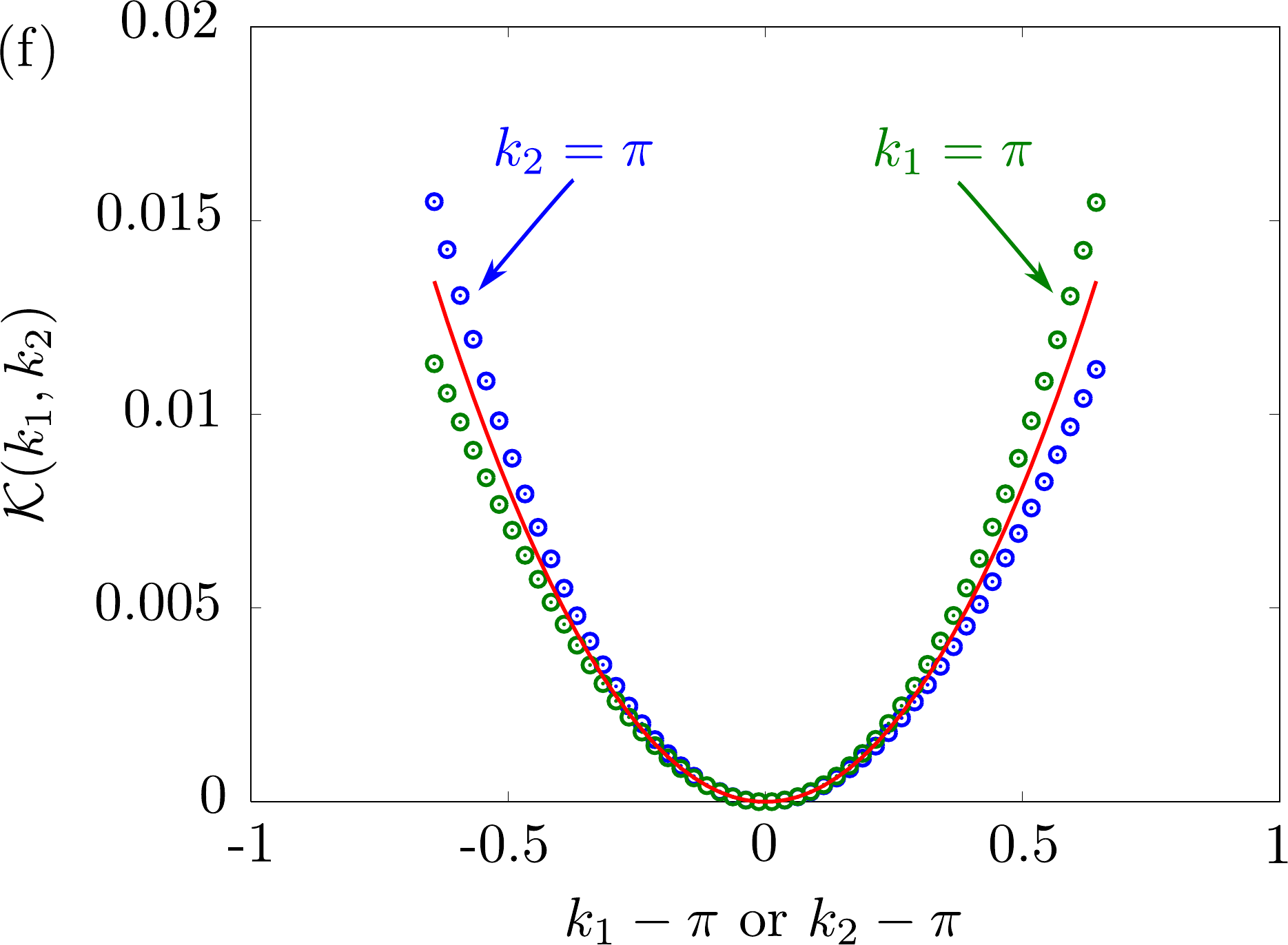}
\caption{(Color online) The RSQA $\mathcal{K}(k_1,k_2)$ is shown in panels (a) and (b) for the BHZ model, in panels (d) and (e) for the KM model as obtained within exact diagonalization on nanoribbons. Parameters for BHZ model: $A=5$, $B=-1$, $M=-2$, $C=D=0$, $L=32$ with 250 discrete $k$ values. Parameters for KM model: $t=1$, $\lambda_{\rm SO}=0.2$, $\lambda_{\rm R,1}=0.05$, $\lambda_{\rm R,2}=0$, $L=32$ unit cells with 250 discrete $k$ values.
In panel (c) one-dimensional cuts are shown for the BHZ model corresponding to $\mathcal{K}(k_1,0)$ (blue circles) and $\mathcal{K}(0,k_2)$ (green circles) as indicated by the black dashed lines in panel (b). The same is repeated in panel (f) for the KM model where $\mathcal{K}(k_1,\pi)$ (blue circles) and $\mathcal{K}(\pi,k_2)$ (green circles) are shown as indicated by the black dashed lines in panel (e).
The red lines in panels (c) and (e) are fits to the low-energy behavior \eqref{rsqa}.}
\label{fig:overap_cylinder_gaughe_inv}
\end{figure*}

We obtain the projected energy spectrum in the effectively one-dimensional Brillouin zone as shown in Fig.\,\ref{fig:bhz-spec}. Now we investigate the eigenstates which correspond to the helical edge states, and which are well separated from the bulk bands, see the green (red) dots associated with momentum $k_1$ ($k_2$) in Fig.\,\ref{fig:bhz-spec} which correspond to the right (left) mover.
Using Eq.\,\eqref{rsqa} we compute the RSQA which has the following characteristic features: (i) for $k_1=k_2$ the RSQA will vanish as both eigenstates were obtained within the same diagonalization process and are, hence, by construction orthogonal; (ii) for $k_1=-k_2$ the RSQA must vanish, too, in order to satisfy the Kramers theorem; (iii) for small momenta $k_1$ and $k_2$ we expect the the RSQA to behave like $\mathcal{K}(k_1,k_2) \approx  k_0^{-2}|k_1^2-k_2^2|$; (iv) if spin-mixing terms $\Delta$ or $\lambda_{\rm R,1/2}$, respectively, are absent the RSQA will identically vanish.
For the BHZ model we show the RSQA in Fig.\,\ref{fig:overap_cylinder_gaughe_inv}\,(a) and (b). It fulfills the properties (i)--(iv). In addition, we show in panel (c) a fit of the $k_1\equiv 0$ and $k_2 \equiv 0$ lines to the low-energy behavior of Eq.\,\eqref{rsqa} which allows us to extract $k_0^{-2}$.

The same strategy is applied for the KM model whose typical spectrum is shown in Fig.\,\ref{fig:kmr-spec} where particle-hole symmetry is broken due to the presence of Rashba spin-orbit coupling. This is also reflected in a slightly asymmetric behavior of the RSQA, see Fig.\,\ref{fig:overap_cylinder_gaughe_inv}\,(d) and (e). Also the $k_1\equiv \pi$ and $k_2 \equiv \pi$ lines shown in panel (f) violate the low-energy behavior property of Eq.\,\eqref{rsqa} for larger $k$ values.

\subsection{Results}

We extract the RSQA amplitude $k_0^{-2}$ by fitting $\mathcal{K}(k_1,k_2)$ to the low-energy  prediction \eqref{rsqa}. Here we followed two strategies: (i) For fixed $\bar{k}_1$ ($\bar{k}_2$) we perform one-dimensional fits along vertical (horizontal) lines in $\mathcal{K}(\bar{k}_1,k_2)$ ($\mathcal{K}(k_1,\bar{k}_2)$) with $k_0^{-2}$ as the only fitting parameter. After doing this for various values of $\bar{k}_1$ and $\bar{k}_2$ we average over all extracted $k_0^{-2}$. (ii) We perform a full two-dimensional fit for $\mathcal{K}(k_1,k_2)$ and extracted $k_0^{-2}$ directly. Both strategies agree in high accuracy. Below we show the behavior of $k_0^{-2}$ as a function of $\Delta$ in case of the BHZ model, and as a function of $\lambda_{\rm SO}$ and $\lambda_{\rm R,1}$ in case of the KM model. It seems that the RSQA amplitude depends linear on $|\Delta|$,
\begin{equation}
k_0^{-2} \approx C_{\rm BHZ} \,\,|\Delta|\ .
\end{equation}
For the KM model the dependence seems to be linear in the Rashba spin-orbit coupling, but proportional to $|\lambda_{\rm SO}|^{-1/2}$ for the intrinsic spin-orbit coupling,
\begin{equation}\label{k0-scaling}
k_0^{-2} \approx C_{\rm KM} \,\frac{|\lambda_{\rm R}|}{\sqrt{|\lambda_{\rm SO}|}}\ .
\end{equation}
$C_{\rm BHZ}$ and $C_{\rm KM}$ are constants. We verified these findings for different widths $L$ of the nanoribbon ruling out finite size effects.

\begin{figure}[b!]
\centering
\includegraphics[scale=0.45]{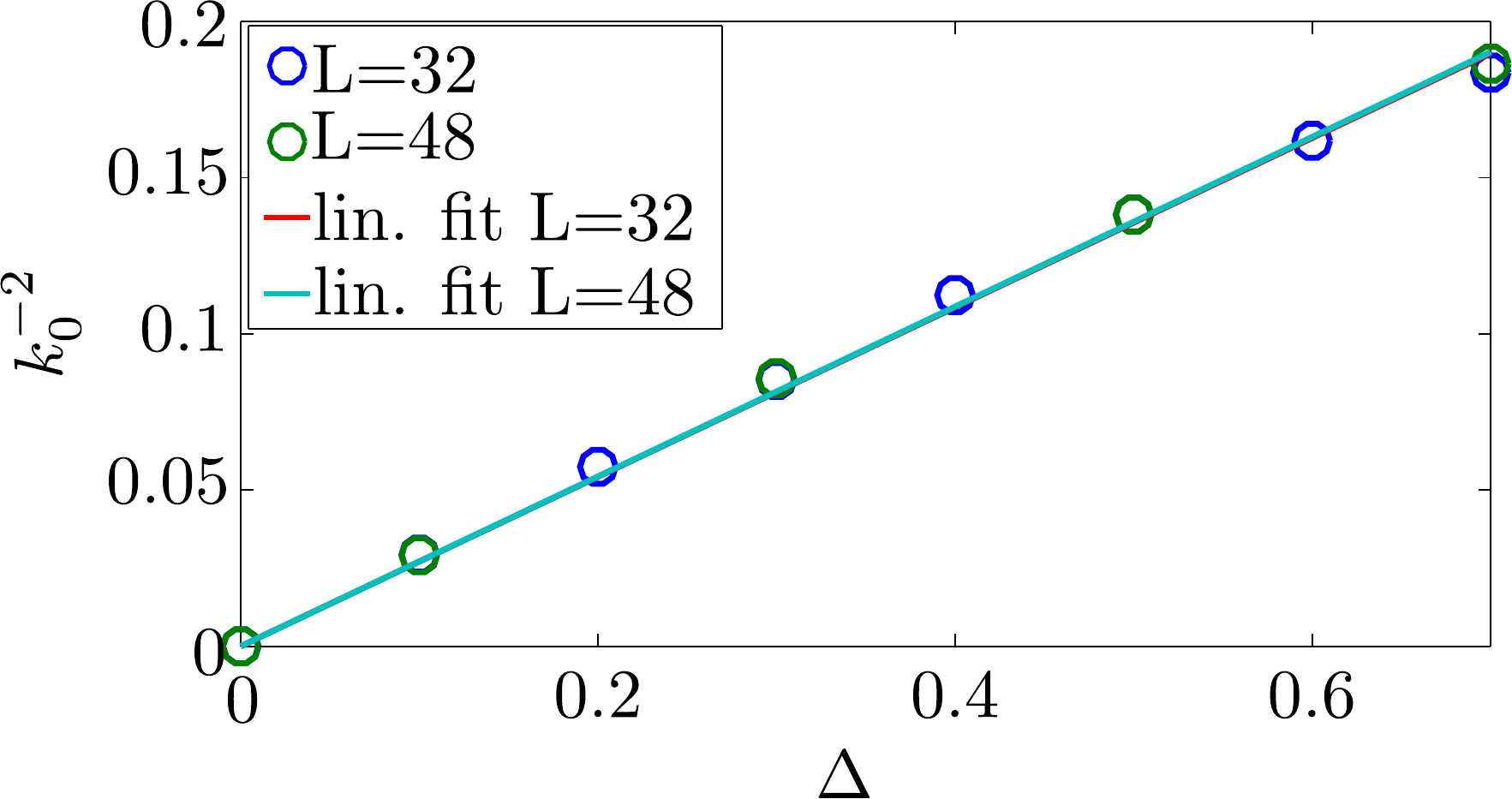}
\caption{RSQA $k_0^{-2}$ as a function of BIA parameter $\Delta$ in the BHZ model. The linear curves fit the data points very well. Parameters are the same as in Fig.\,\ref{fig:overap_cylinder_gaughe_inv}.}
\label{fig:k0-fit-bhz}
\end{figure}

\begin{figure}[t]
\includegraphics[width=0.7\linewidth]{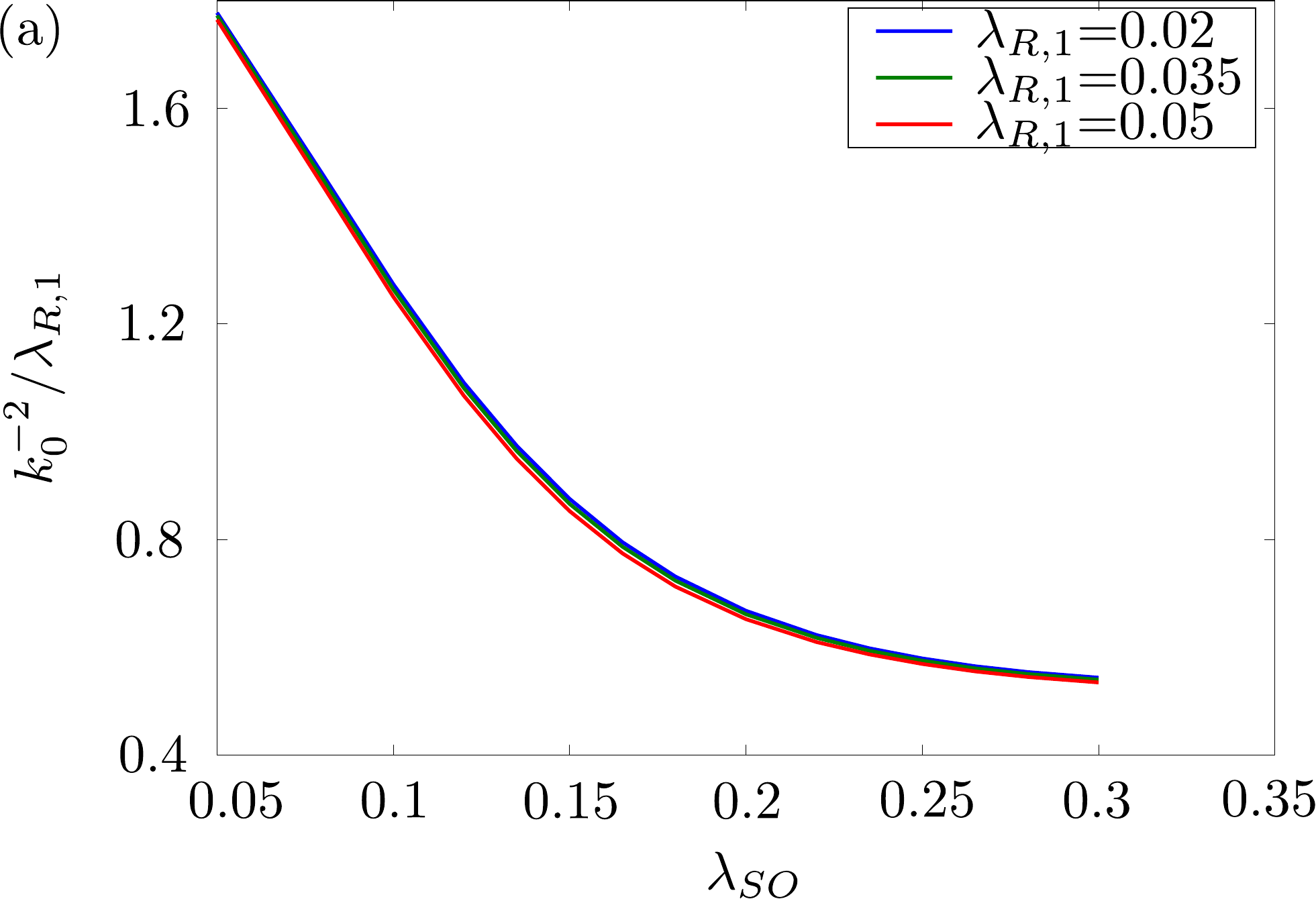}
\includegraphics[width=0.7\linewidth]{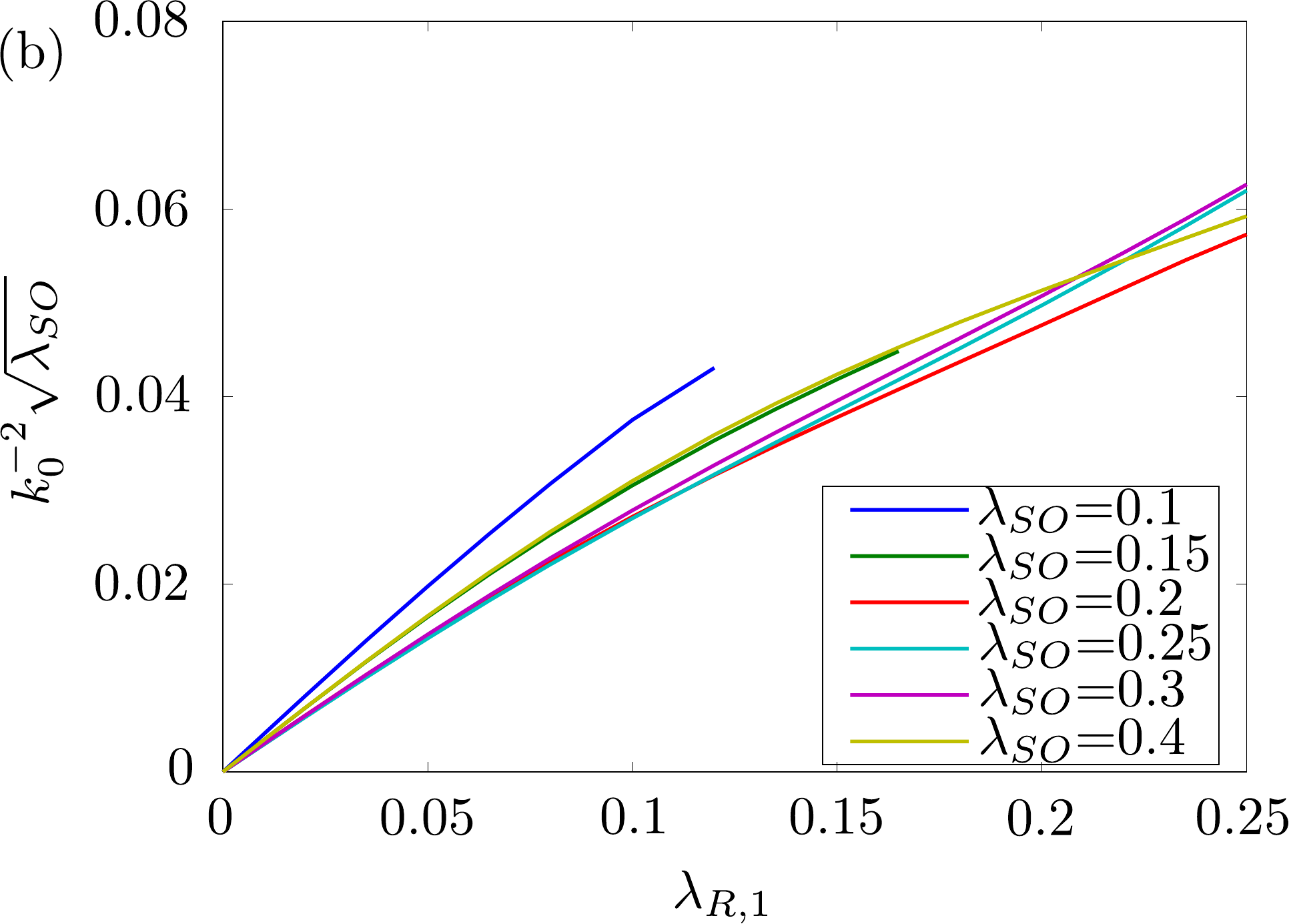}
\caption{(Color online) Parameter dependence of $k_0^{-2}$ in the KM model (a) with respect to $\lambda_{\rm SO}$  and (b) with respect to $\lambda_{\rm R,1}$. The $y$ axes have been rescaled according to Eq.\,\eqref{k0-scaling}. See the main text for details. Parameters used correspond to those in Fig.\,\ref{fig:overap_cylinder_gaughe_inv}.}
\label{fig:evolution_k0}
\end{figure}

\begin{figure}[t]
\includegraphics[width=0.9\linewidth]{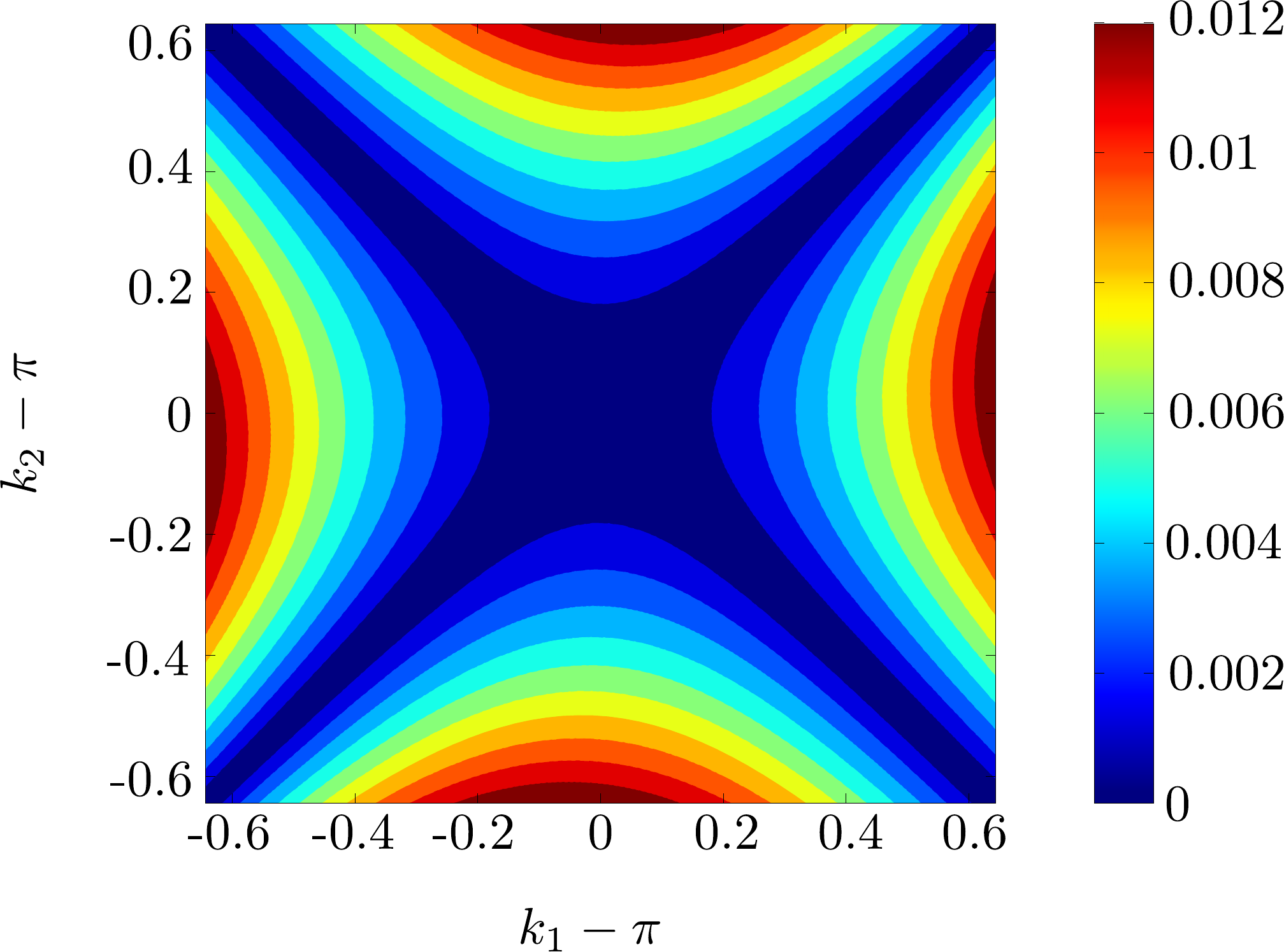}
\caption{(Color online) The RSQA $\mathcal{K}(k_1,k_2)$ for the KM model with second Rashba term as present in Stanene and similar materials. Parameters for KM model: $t=1$, $\lambda_{\rm SO}=0.2$, $\lambda_{\rm R,1}=0$, $\lambda_{\rm R,2}=0.5$, $L=32$ unit cells with 250 discrete $k$ values.}
\label{fig:overlap_silicene}
\end{figure}

We also consider the KM model with the second Rashba term.
That allows a description of the proposed TI phase in silicene,\cite{mahatha14,vogt12,ezawa-12prl055502,rachel-14prb195303} germanene, and stanene.\cite{yong13} In Fig.\,\ref{fig:overlap_silicene} we show a representative example of RSQA in the presence of the second Rashba term as realized \eg in stanene (neglecting the first Rashba term). We emphasize that treating both Rashba terms hardly influences the results as $\lambda_{\rm R,1}$ is much smaller than $\lambda_{\rm R,2}$ in these materials.\cite{ezawa-12prl055502} We further observe that the presence of $\lambda_{\rm R,2}$ does not break particle-hole symmetry;  we find, however, that the second Rashba term is a much weaker source for generating RSQA. Fig.\,\ref{fig:overlap_silicene} is computed for $\lambda_{\rm R,2}=0.5$ which is ten times larger than the value of $\lambda_{\rm R,1}$ in Fig.\,\ref{fig:overap_cylinder_gaughe_inv}. In silicene, germanene, and stanene, $\lambda_{\rm R,1}$ still is so small that we can safely neglect it.

Using realistic system parameters for Silicene, Germanene, and Stanene as obtained within {\it ab inito} calculations\,\cite{liu11PRB84_195430}
we find material values for $k_0^{-2}$ as presented in Table~\ref{table-k0-realistic}. We also computed $k_0^{-2}$ based on {\it ab inito} calculations for a closely related band structure which was proposed to describe a possible TI phase in Na$_2$IrO$_3$.\cite{shitade-09prl256403} Although recent experiments suggest that the true material is due to strong Coulomb interactions in a magnetically ordered Mott phase\,\cite{singh-10prb064412}  the non-interacting band structure provides a bond-dependent generalization of the KM model. This sodium-iridate band structure crucially differs, however, from the KM case as the intrinsic spin-orbit coupling genuinely breaks the axial spin symmetry. Considering the large values we find for $k_0^{-2}$ makes possible TI phases in transition metal oxides an promising platform to detect large corrections to the conductivity due to rotation of the spin quantization axis.

\begin{table}[b!]
\begin{tabular}{c|c|c}
& $~k_0^{-2} {\tilde a}^{-2}$~ & ~ $k_0 \tilde a$~  \\[1pt] \hline\hline
  Silicene & $~~2.4341\cdot10^{-7}~~$ & ~~$2027$~~\\
  ~Germanene~~ & $6.8594\cdot10^{-5}$ & $121$ \\
  Stanene & $1.0282\cdot10^{-4}$ &$99$ \\
  Na$_2$IrO$_3$ & $2.6503\cdot10^{-2}$ & 6.14 
\end{tabular}
\caption{Values of $k_0^{-2}$ and $k_0$ multiplied with the $\tilde a$, the distance between unit cells, for the buckled honeycomb lattice materials Silicene, Germanene, and Stanene as well as for the transition metal oxide Na$_2$IrO$_3$. The used system parameters are taken from Refs.\,\onlinecite{liu11PRB84_195430} and \onlinecite{shitade-09prl256403}. The first Rashba-term is absent, $\lambda_{\rm R,1}=0$, and in case of Na$_2$IrO$_3$ none of the Rashba terms is considered.}
\label{table-k0-realistic}
\end{table}

Here we have determined the RSQA on nanoribbons for the non-interacting TI models. An exciting perspective provides the idea to study interacting TI models,\cite{rachel-10prb075106} \ie topological Hubbard-like models, on cylinder geometries.\cite{wu-12prb205102,laubach-14prb165136} A promising model to investigate an interacting generic helical liquid would be the Kane-Mele-Hubbard model in the presence of Rashba spin-orbit coupling.\cite{laubach-14prb165136}

%
%
\section{Rotation of  spin-quantization axis II: continuum disks}

\subsection{Set-up}

In the previous section, we worked out the effect of breaking the axial spin symmetry on the spin texture of the helical edge states. We showed that it results in a rotation of the spin-quantization axis as a function of momentum. In most of the experimental situations relevant for transport, measurement geometries like Hall bars are used. Here momentum is not a good quantum number anymore and spatial symmetries are lost. We approach this situation in two steps: first we try to understand how the breaking of axial spin symmetry is manifested in real space when the topological insulator sample has a circular symmetry, \ie is rotationally invariant. In the next section, we eventually consider real space samples of TIs which neither possess translational nor rotational symmetry.

\begin{figure}[t]
\includegraphics[width=0.5\linewidth]{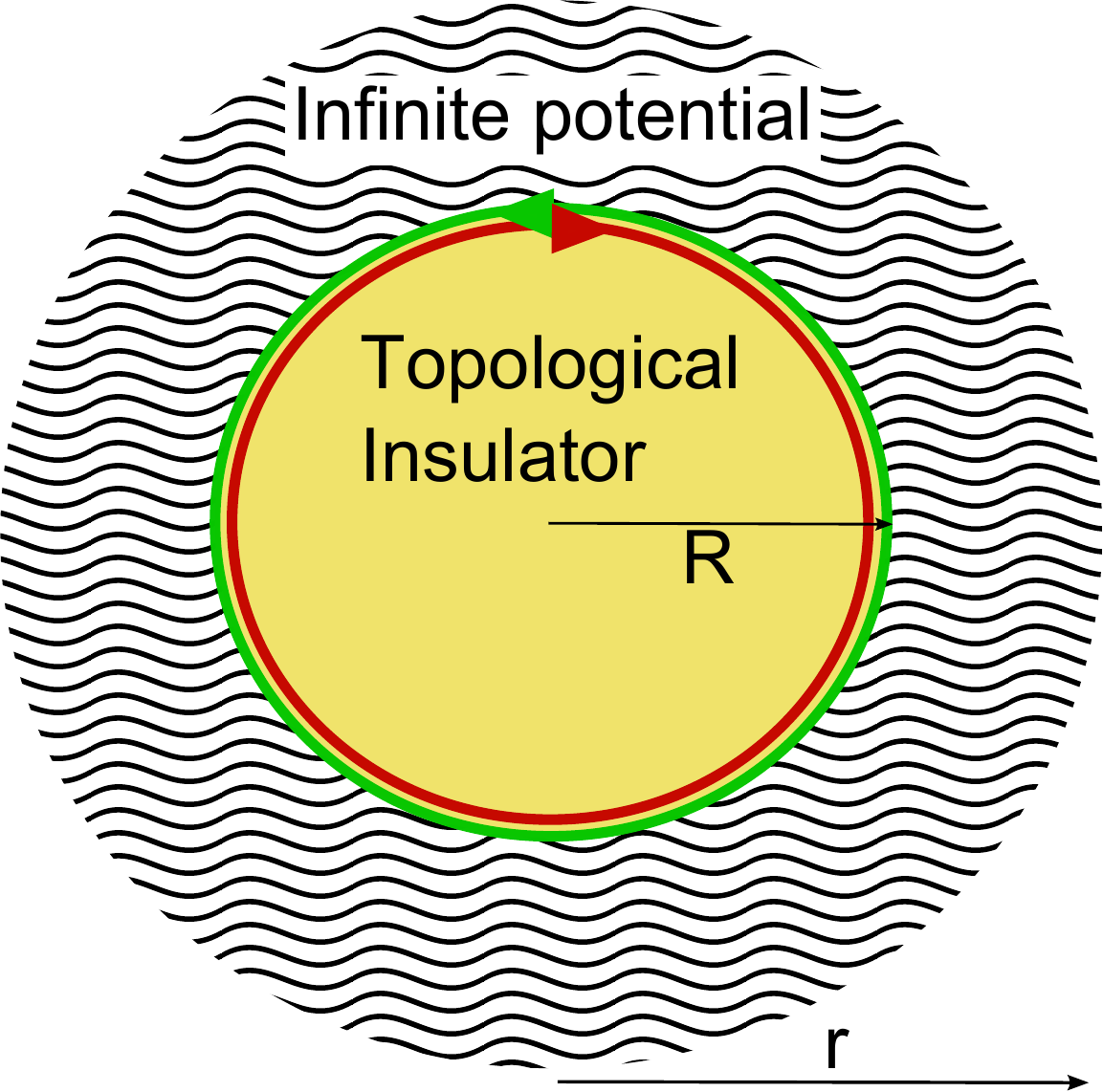}
\caption{Setup as explained in the main text. $R$ denotes the disk radius while $r$ the distance from the origin. The disk is surrounded by an infinite potential. The helical edge modes are located
at the boundary between TI and infinite potential.}
\label{fig:analyt_disk_sketch}
\end{figure}

In this section, we discuss the BHZ model in the presence of BIA for a disk geometry. In order to solve this model analytically, we switch to the continuum description Eq.\,\eqref{eq:BHZ} where the BIA term couples both spin channels.
We assume that the disk is surrounded by an infinite potential in order to guarantee that edge states cannot penetrate into vacuum, see Fig.~\ref{fig:analyt_disk_sketch}.
The edge-states are moving clockwise and counter-clockwise, respectively, at the disk edge at radius $R$.

Continuum models for TIs defined on disk geometries were investigated previously. The BHZ model with a Rashba term was investigated in Ref.~\onlinecite{michetti11PRB83_125420} where the spectrum and wave functions were computed for TI Aharonov--Bohm rings and disks. A TI antidot (one might think of a ``missing disk'') described within the BHZ model without BIA was considered in Ref.~\onlinecite{shan11PRB84_35307}.
In Ref.\,\onlinecite{rakyta-15prb125412} the KM model for silicene in a magnetic field with and without intrinsic spin-orbit coupling $\lambda_{\rm SO}$ was discussed. Ref.~\onlinecite{kundu11} investigated short cylinders made from 3D TIs.

We start from Eq.\,\eqref{eq:BHZ} and set $C=D=0$ as the corresponding terms only shift the total energy and do not influence the spin texture of the helical edge states. The momentum-space representation of \eqref{eq:BHZ} is readily transformed to real space by rewriting
$k_x=-i\partial_x$ and $k_y=-i\partial_y$, respectively. The physical spin is represented by the Pauli matrices $s^\alpha$. In real space, Eq.~\eqref{eq:BHZ} can be written in a compact way as
\begin{eqnarray}
\mathcal{H}&=& \mathcal{H}_{\rm BHZ}+\mathcal{H}_{\rm BIA}\nonumber\\[5pt]
&=&-iA(\partial_x\sigma_x\otimes s_z-\partial_y\sigma_y \otimes \mathbb{I}_{2x2})\\[5pt]
& & {} + \left(M+B\partial_x^2+B\partial_y^2\right)\sigma_z\otimes\mathbb{I}_{2x2}+\Delta s_y \otimes \sigma_y \label{eq:initial_disk_H} \nn
\end{eqnarray}
In order to exploit the rotational symmetry of the system we introduce polar coordinates $x=r\cos(\phi)$, $y=r\sin(\phi)$. The constant term $\mathcal{H}_{\rm BIA}$ is not affected by the transformation of coordinate system. In polar coordinates, $\mathcal{H}_{\rm BHZ}$ reads

\begin{eqnarray}
\mathcal{H}_{\rm BHZ}~&=&~-A\Big( e^{i\phi\sigma_z\otimes s_z}i\partial_r\sigma_x\otimes s_z \Big.\nn\\[5pt]
&&\Big.+e^{i\phi\sigma_z\otimes s_z}\frac{l_z}{r}\sigma_y\otimes \mathbb{I}_{2x2}\Big) \\[5pt]
&&+\left(M+B\partial_r^2-B\frac{l_z^2}{r^2}+B\frac{\partial_r}{r}\right)\sigma_z\otimes \mathbb{I}_{2x2}\ ,\nn
\end{eqnarray}
where $l_z=-i\partial_\phi$ denotes the orbital angular momentum. We further define the {\it total angular momentum} $j_z$ as
\begin{equation}
j_z=l_z-\frac{1}{2}(s_z\sigma_z).
\end{equation}
This operator commutes with the Hamiltonian, $[\mathcal{H},j_z]=0$, 
so the eigenstates of the Hamiltonian can be chosen as eigenstates of $j_z$,
\begin{equation}
j_z\Psi_j(r,\phi)=j\Psi_j(r,\phi)\ .
\end{equation}
By solving this equation the angular part can be determined:
\begin{equation}
\begin{pmatrix}
\Psi_{jE_1+}(r,\phi)\\[3pt]
\Psi_{jH_1+}(r,\phi)\\[3pt]
\Psi_{jE_1-}(r,\phi)\\[3pt]
\Psi_{jH_1-}(r,\phi)\\
\end{pmatrix}=e^{ij\phi}\begin{pmatrix}
\Psi_{jE_1+}(r)e^{i\phi/2}\\[3pt]
\Psi_{jH_1+}(r)e^{-i\phi/2}\\[3pt]
\Psi_{jE_1-}(r)e^{-i\phi/2}\\[3pt]
\Psi_{jH_1-}(r)e^{i\phi/2}\\[0pt]
\end{pmatrix}.
\end{equation}
\phantom{x x x}\\
Moreover, the $\phi$ periodicity of the wave function, $\Psi_j(r,\phi)=\Psi_j(r,\phi+2\pi)$, imposes the constraint $j+1/2\in\mathbb{Z}$. Thus, we have completely determined the angular part of the wave function.

In order to solve the radial part of the wave function we apply the Hamiltonian $\mathcal{H}$ to the wave function, $\mathcal{H}\Psi_j(r,\phi)=E\Psi_j(r,\phi)$, where $E$ denotes the corresponding energy. This yields the following set of equations:
\begin{widetext}
\begin{subequations}
\begin{eqnarray}
\left(M+B\partial_r^2-B\frac{(j+1/2)^2}{r^2}+B\frac{1}{r}\partial_r-E\right)\Psi_{jE_1+}(r) + Ai\left(-\partial_r+\frac{j-1/2}{r}\right)\Psi_{jH_1+}(r)  -\Delta\Psi_{jH_1-}(r)&=&0\\[0.3em]
Ai\left(-\partial_r-\frac{j+1/2}{r}\right)\Psi_{jE_1+}(r) +\left(-M-B\partial_r^2+B\frac{(j-1/2)^2}{r^2}-B\frac{1}{r}\partial_r-E\right)\Psi_{jH_1+}(r) + \Delta \Psi_{jE_1-}(r)&=&0\\[0.3em]
\Delta \Psi_{jH_1+}(r) + \left(M+B\partial_r^2-B\frac{(j-1/2)^2}{r^2}+B\frac{1}{r}\partial_r-E\right)\Psi_{jE_1-}(r) + Ai\left(\partial_r+\frac{j+1/2}{r}\right)\Psi_{jH_1-}(r)&=&0\\[0.3em]
-\Delta\Psi_{jH_1-}(r) + Ai\left(\partial_r-\frac{j-1/2}{r}\right)\Psi_{jE_1-}(r) + (-M-B\partial_r^2+B\frac{(j+1/2)^2}{r^2}-B\frac{1}{r}\partial_r-E)\Psi_{jH_1-}(r)&=&0.
\end{eqnarray}
\end{subequations}
\end{widetext}
We assume that each component of the radial part of $\Psi_j(r,\phi)$ can be written as
\begin{equation}
\Psi_{j\alpha}(r)=\Psi_{j\alpha}Z_j(\sqrt{p}r)
\end{equation}
where $Z_j$ is a Bessel function (either of the first kind, second kind, or a Hankel function, depending on the boundary conditions) and $\alpha$ labels the orbital and spin degrees of freedom, $\alpha=(E/H,\pm)$. In the following we will often make use of the recursion relation for Bessel functions,
\begin{equation}\label{bessel-recursion}
\sqrt{p}Z_{j\pm1}(\sqrt{p}r)=\frac{j}{r}Z_j(\sqrt{p}r)\mp\partial_rZ_j(\sqrt{p}r)\ .
\end{equation}
In addition we rewrite the terms proportional to $B$ as follows,

\begin{equation}\label{identity01}
\left(\partial_r^2-\frac{j^2}{r^2}+\frac{1}{r}\partial_r\right)=\left(\partial_r+\frac{j+1}{r}\right)\left(\partial_r-\frac{j}{r}\right)\ .
\end{equation}
Applying Eq.~\eqref{identity01} to the Bessel function and using the recursion formula \eqref{bessel-recursion} yields

\begin{equation}
\left(\partial_r^2-\frac{j^2}{r^2}+\frac{1}{r}\partial_r\right)Z_j(\sqrt{p}r)\nonumber=-pZ_j(\sqrt{p}r)\ . \nn
\end{equation}
This relation enables us to eliminate all derivatives from the set of coupled equations and to simplify them.
These equations are linear in the $\Psi$'s which allows us to organize them in a matrix equation.
Thus we can write them compactly as
\begin{widetext}
\begin{equation}
\begin{pmatrix}
M-Bp-E&Ai\sqrt{p}& 0 & -\Delta\\[5pt]
-Ai\sqrt{p} &-M+Bp-E & \Delta & 0\\[5pt]
0 & \Delta &M-Bp-E&Ai\sqrt{p}\\[5pt]
-\Delta & 0&-Ai\sqrt{p}&-M+Bp-E\\
\end{pmatrix}
\begin{pmatrix}
\Psi_{jE_1+}\\[5pt]
\Psi_{jH_1+}\\[5pt]
\Psi_{jE_1-}\\[5pt]
\Psi_{jH_1-}\\
\end{pmatrix}
\equiv \tilde{\mathcal{H}} \vec \Psi
=0.
\end{equation}
\end{widetext}
To find nontrivial solutions, we require vanishing determinant of $\tilde{\mathcal{H}}$,
\begin{equation}
\left[(M-Bp)^2-E^2+\Delta^2+A^2p\right]^2-4\Delta^2A^2p=0\ .
\end{equation}
Solving this polynomial equation with respect to $p$, we obtain four roots $p_n$, $n=1,2,3,4$ (the roots become doubly degenerate if $\Delta= 0$ since $\tilde{\mathcal{H}}$ is block-diagonal in this case).
It is possible to find closed expressions for the $p_n$ for arbitrary parameters $A$, $B$, $M$, $\Delta$, and the energy $E$. Since these expressions are very long, we omit them here. Nonetheless we always work with the exact expressions for $p_n$.

The choice of Bessel function is fixed by the requirements not to diverge at the origin $r=0$, thus we work with Bessel functions of the first kind, denoted  as $J_j(\sqrt{p}r)$.
Now we can write the eigenstates in the presence of finite $\Delta$ as

\begin{eqnarray}
&&\Psi_j(r,\phi)= \\[10pt]
&&\sum_{n=1}^4\!A_ne^{\!ij\phi}\!\begin{pmatrix}
i\frac{(M-Bp_n)^2-E^2+\Delta^2-A^2p_n}{2A\sqrt{p_n}(M-Bp_n-E)}J_{j+\frac{1}{2}}(\sqrt{p_n}r)e^{i\phi/2}\\[10pt]
J_{j-\frac{1}{2}}(\sqrt{p_n}r)e^{-i\phi/2}\\[10pt]
\frac{(M-Bp_n)^2-E^2-\Delta^2+A^2p_n}{2\Delta(M-Bp_n-E)}J_{j-\frac{1}{2}}(\sqrt{p_n}r)e^{-i\phi/2}\\[10pt]
i\frac{(M-Bp_n)^2-E^2+\Delta^2+A^2p_n}{2\Delta A\sqrt{p_n}}J_{j+\frac{1}{2}}(\sqrt{p_n}r)e^{i\phi/2}
\end{pmatrix}.\nonumber\\[-3pt]
\nonumber\label{eq:analyt_wave_func}
\end{eqnarray}
The paramters $A$, $B$, $M$, and $\Delta$ are chosen such that we stay in the topological phase.
Now we force the energy $E$ to be smaller than the band gap which will allow us to compute the helical edge states.
In this parameter regime we find  the four roots $p_n$ to be complex. We impose boundary conditions such that the wave function vanishes at the boundary of the disk, $\Psi_{j}(r=R,\phi)=0$.
Inserting this condition for the wave functions into the previous equations leads to the requirement that the coefficient matrix has a vanishing determinant:
\begin{figure*}[t]
\includegraphics[width=0.35\textwidth]{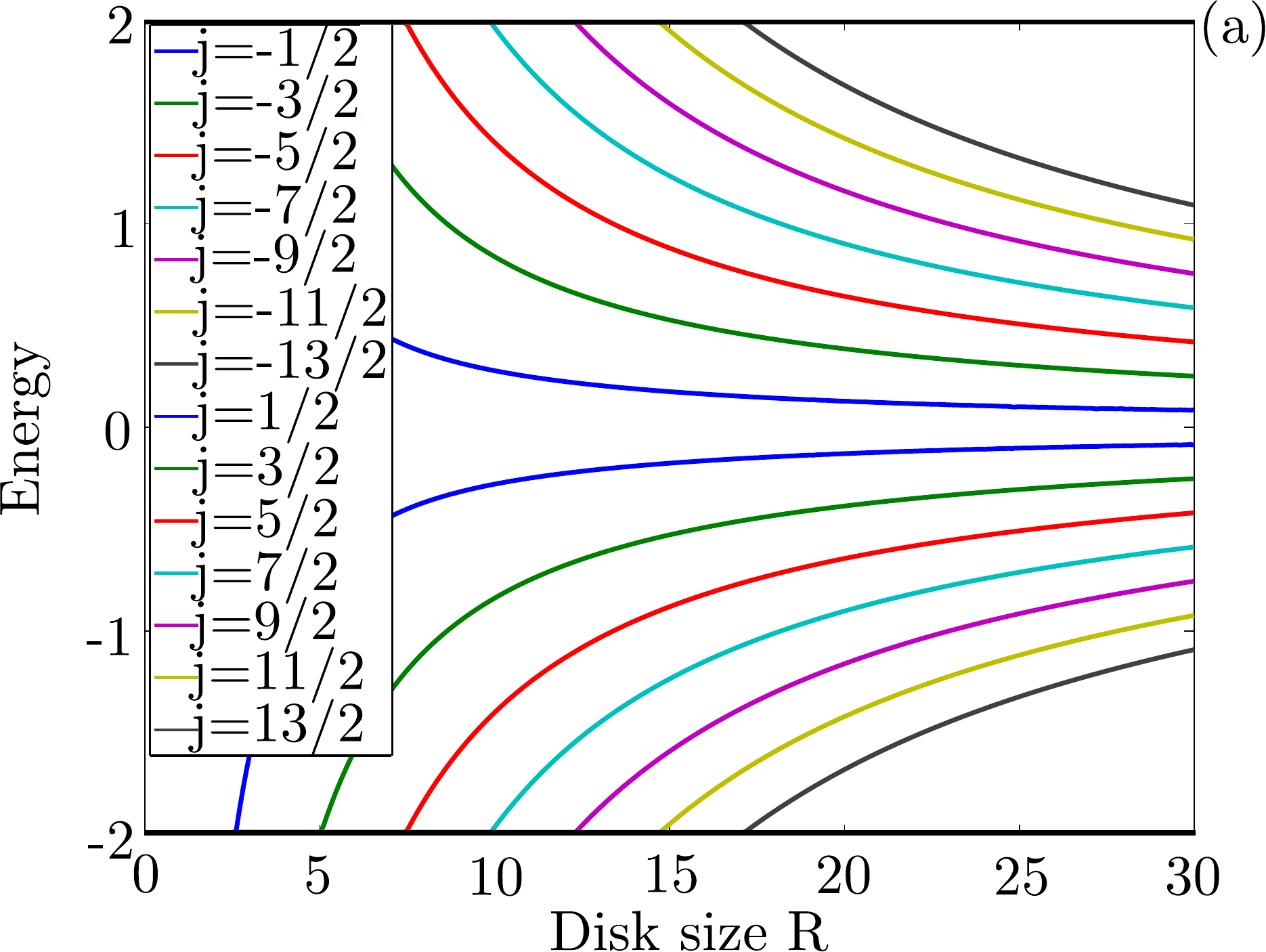}\qquad
\includegraphics[width=0.37\textwidth]{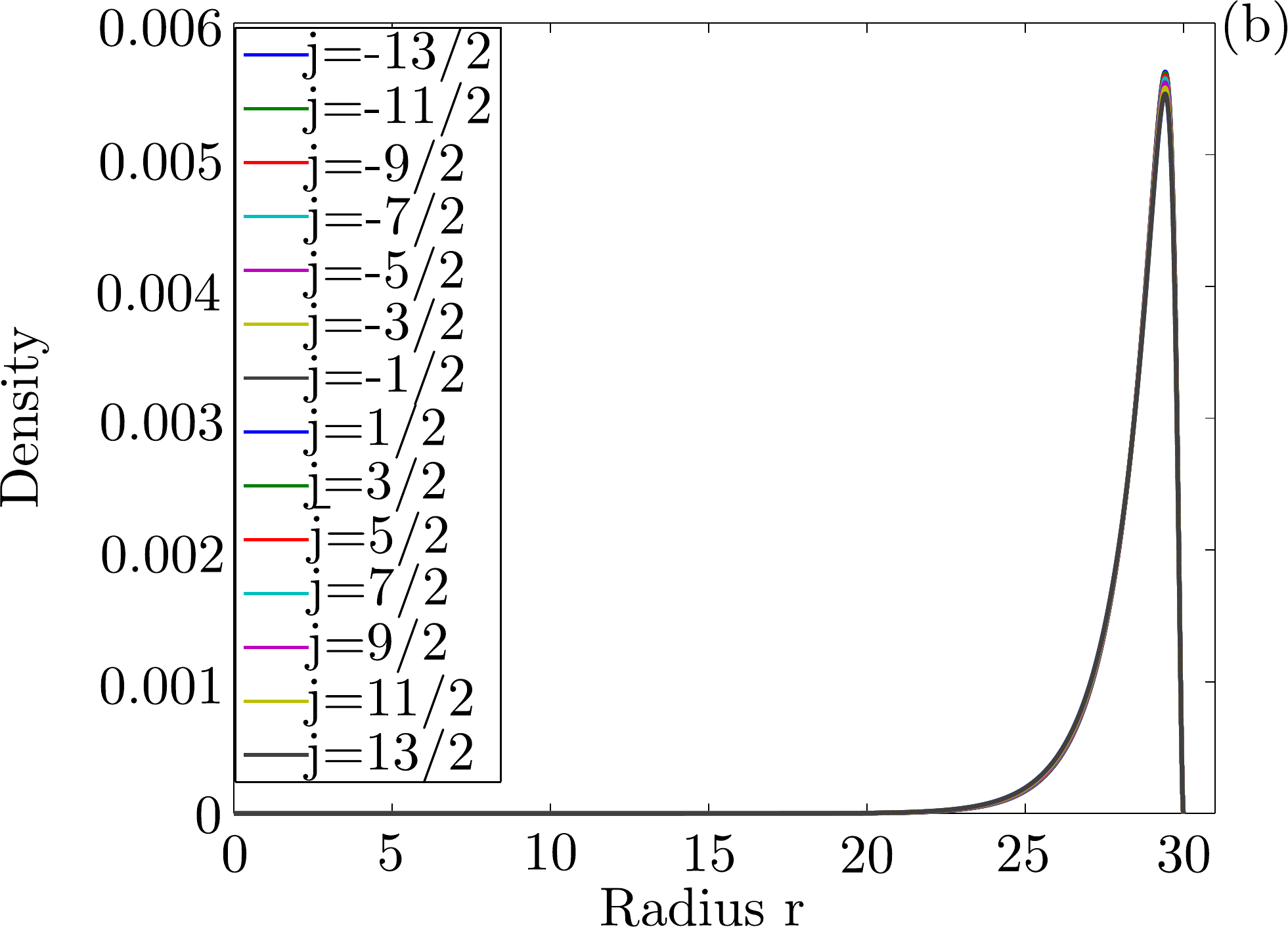}\\
\hspace{-10pt}\includegraphics[width=0.37\textwidth]{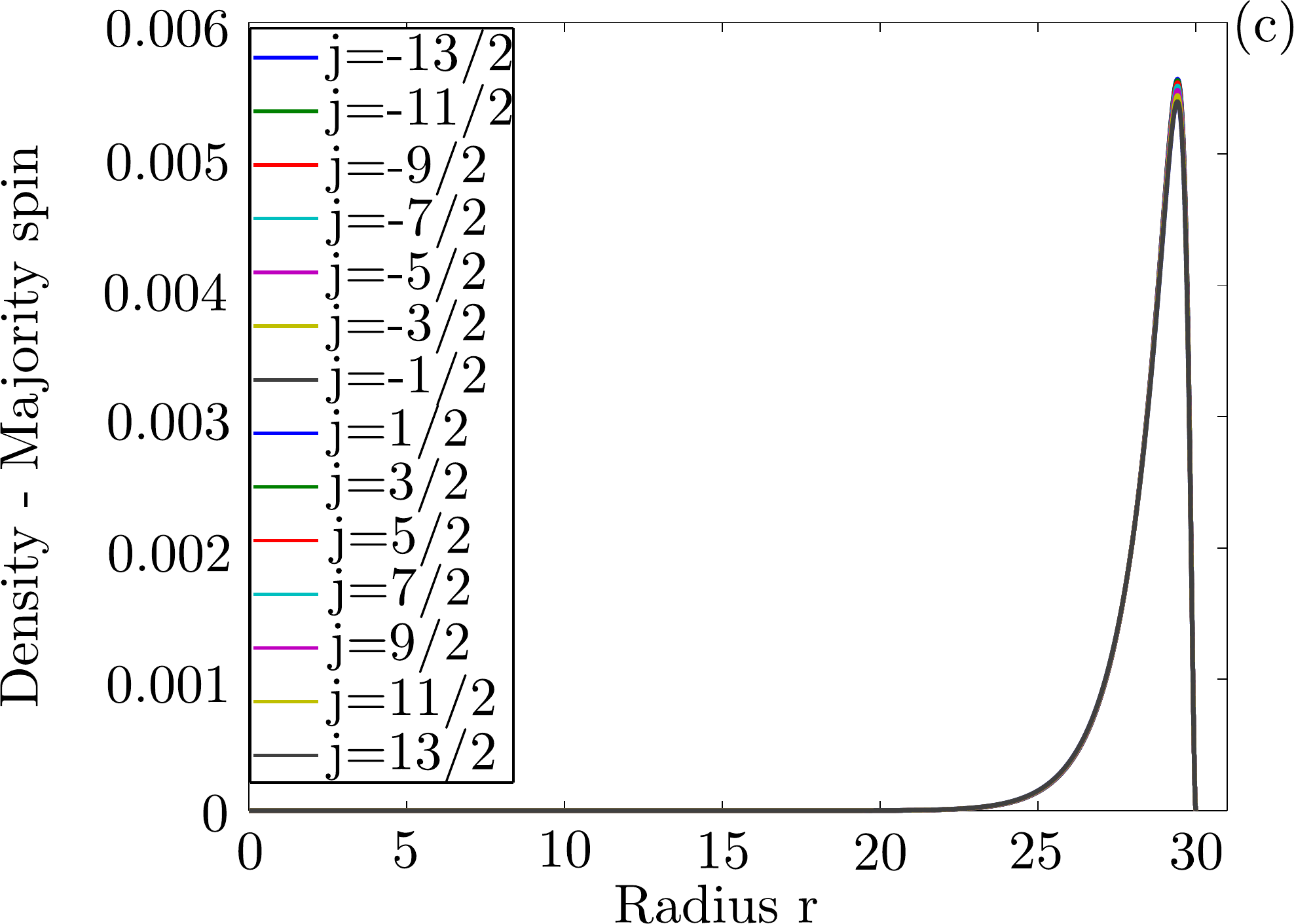}\qquad
\includegraphics[width=0.37\textwidth]{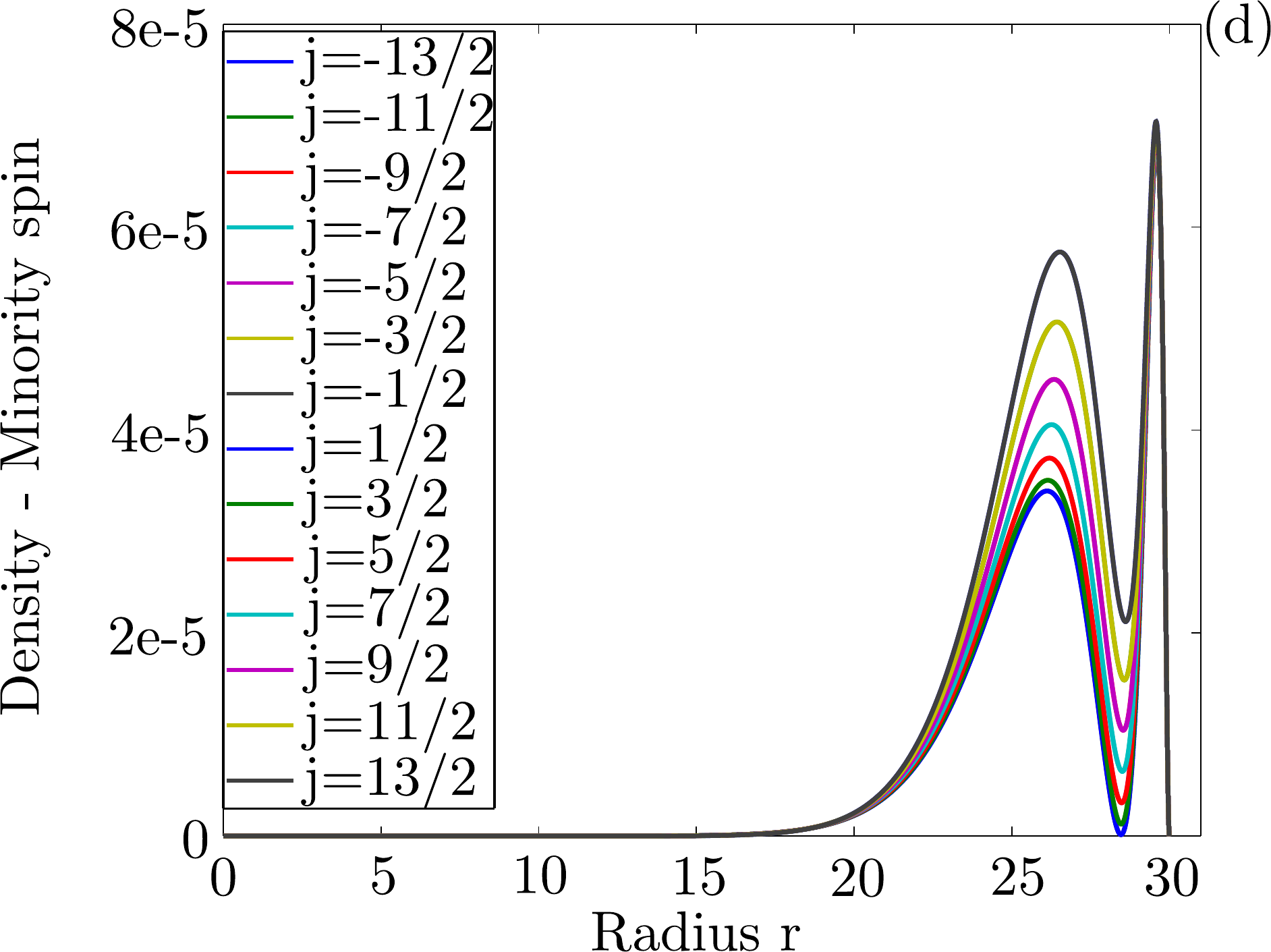}
\caption{(Color online) (a) Energy spectrum as a function of the disk size $R$. (b) Total radial density $|\Psi_j(r,\phi=0)|^2$ as a function of the radius $r$ for a disk of size $R=30$. In (c) and (d) the radial density is again shown for the different spin channels separately. Parameters used: $A=5$, $B=-1$, $M=-2$, $\Delta=0.5$. Note the different scale in (d) compared to (c).}
\label{fig:analyt_energy_wavefunc}
\end{figure*}

\begin{widetext}
\begin{equation}\label{final-det}
\det\begin{pmatrix}
~~\zeta_1 J_{j+1/2}(\sqrt{p_1}R) ~&~ \zeta_2 J_{j+1/2}(\sqrt{p_2}R) ~&~\zeta_3 J_{j+1/2}(\sqrt{p_3}R) ~&~\zeta_4 J_{j+1/2}(\sqrt{p_4}R)~~\\[5pt]
J_{j-1/2}(\sqrt{p_1}R) &  J_{j-1/2}(\sqrt{p_2}R) & J_{j-1/2}(\sqrt{p_3}R) & J_{j-1/2}(\sqrt{p_4}R)\\[5pt]
\eta_1 J_{j-1/2}(\sqrt{p_1}R) & \eta_2 J_{j-1/2}(\sqrt{p_2}R) &\eta_3 J_{j-1/2}(\sqrt{p_3}R) & \eta_4 J_{j-1/2}(\sqrt{p_4}R)\\[5pt]
\xi_1 J_{j+1/2}(\sqrt{p_1}R) &\xi_2 J_{j+1/2}(\sqrt{p_2}R) & \xi_3 J_{j+1/2}(\sqrt{p_3}R) &\xi_4 J_{j+1/2}(\sqrt{p_4}R)\\
\end{pmatrix}=0\ .
\end{equation}
\end{widetext}
where we introduced,
\begin{equation}
\zeta_n=\frac{(M-Bp_n)^2-E^2+\Delta^2-A^2p_n}{\sqrt{p_n}(M-Bp_n-E)}\ ,
\end{equation}
\begin{equation}
\xi_n=\frac{(M-Bp_n)^2-E^2+\Delta^2+A^2p_n}{\sqrt{p_n}}\ ,
\end{equation}
\begin{equation}
\eta_n=\frac{(M-Bp_n)^2-E^2-\Delta^2+A^2p_n}{(M-Bp_n-E)}\ .
\end{equation}
By inserting the parameters $A$, $B$, $M$, $\Delta$, $j$, and $R$ into the determinant, Eq.\,\eqref{final-det} only has a solution for two different values of the energy, $+E$ and $-E$,  to be determined. Once we have found the energy of the edge state, inserting all parameters and the energy $E$ into the wavefunction renders the corresponding eigenstates. The larger the radius $R$ is the more energy levels with higher values of total angular momentum $j$ fit into the band gap.

We compute energies and eigenstates for various disk sizes $R$ leading to the spectral plots Fig.~\ref{fig:analyt_energy_wavefunc}\,(a). Note that the energy spectrum is doubly degenerate as expected for a helical liquid (Kramers theorem). In Fig.~\ref{fig:analyt_energy_wavefunc}\,(a), energy levels for the clockwise moving edge state  are labeled from $j=13/2$ (level with lowest energy) to $j=-13/2$ (level with highest energy). For the degenerate counter-clockwise moving edge state the $j$ labels have opposite sign (note that for the sake of clarity, the legend has been omitted for these levels). All energy levels are fully classified by energy and  total angular momentum.
At sufficiently large $R$ the energy levels tend to be equally spaced.
When $R\rightarrow0$, there are no energy levels which remain in the gap. We need a minimal disk radius to obtain edge states. This behavior seems intuitively correct as helical edge states possess a finite penetration length; very short radii allow the clockwise and counter-clockwise moving modes at ``opposite edges'' of the disk to overlap and gap out.

We computed the energy spectrum for different values of $\Delta$. Larger values of $\Delta$ result in an energy spectrum with smaller level spacing; the energy levels also appear now for smaller radii.
Note that the particle hole symmetry remains intact at all times (independent of the choices for $R$ and $\Delta$).
As a consistency check, we recover for $\Delta\rightarrow0$ the results for the disk without BIA (\ie for two uncoupled Chern insulators with opposite chirality).

Now we will discuss the properties of the eigenstates. The radial density of the wave function $|\Psi_j(r,\phi=0)|^2$ can be obtained from Eq.~(\ref{eq:analyt_wave_func}) where $\phi$ can be set to $0$ due to rotational symmetry. The values of $A_n$ are determined from the normalization condition. In addition, they need to fulfill that $\Psi_j(r=R,\phi)=0$. The resulting total radial density is shown in Fig.~\ref{fig:analyt_energy_wavefunc}(b). In panels (c) and (d) we have shown the spin-resolved radial densities. The shape of the total radial density is determined by the one of the majority spin. The contribution of the minority spin is smaller, but it penetrates more deeply in the topological insulator disk and the structure is very different compared to the majority spin. But the resulting broadening of the total density is very small. Note that we reproduce the behavior of the minority spin (Fig.\,\ref{fig:analyt_energy_wavefunc}\,(d)) for the tight binding disks discussed in the following section and it is, hence, not an artifact of the continuum approximation.

\begin{figure}[b!]
\includegraphics[width=0.42\textwidth]{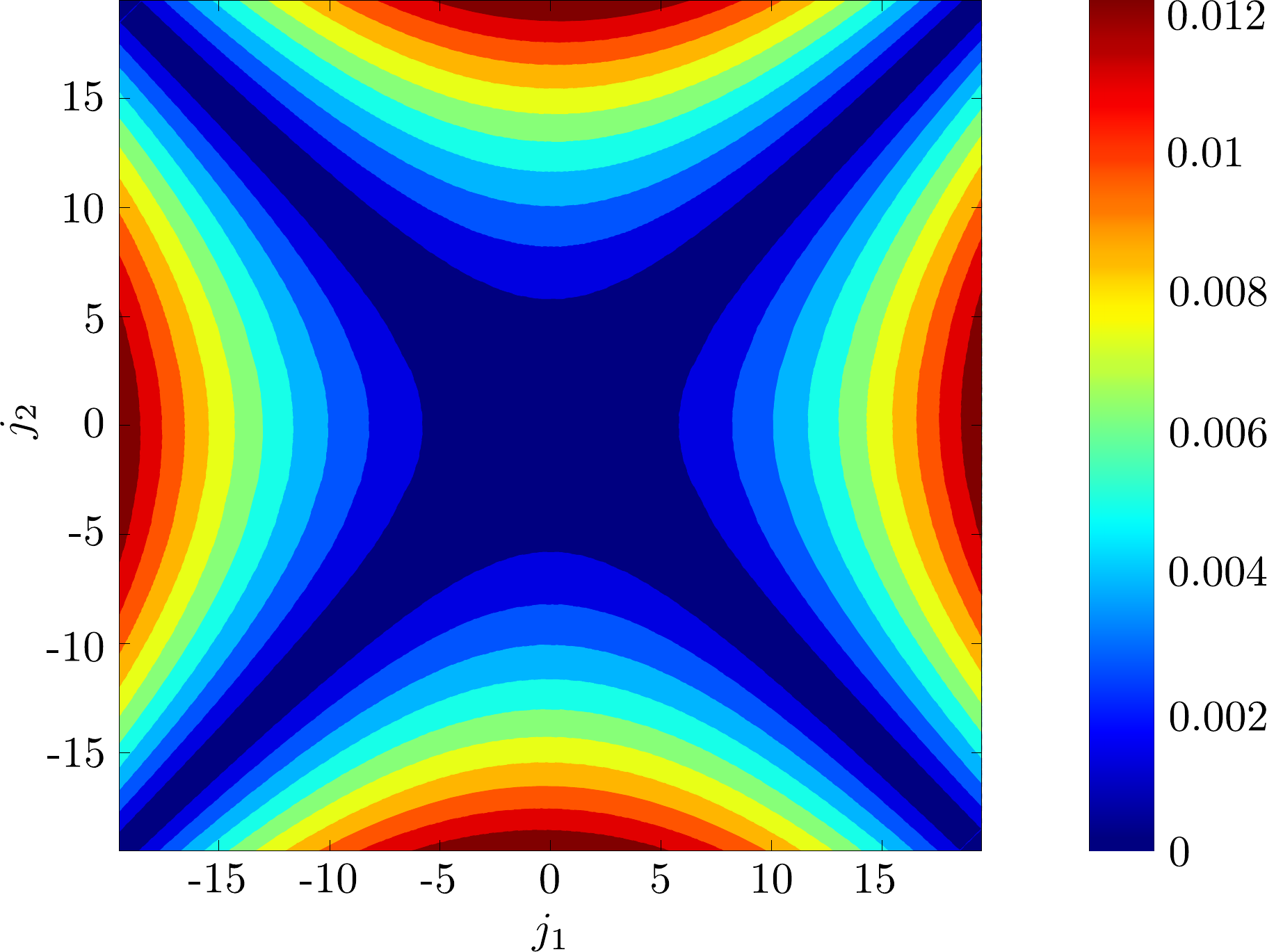}
\caption{A representative example for RSQA $\mathcal{K}(j_1,j_2)$ as a function of total angular momenta $j_1$ and $j_2$ of the continuum disk is shown. Parameters used: $A=5$, $B=-1$, $M=-2$, $\Delta=0.5$, and  disk size  $R=50$.}
\label{fig:analyt_overlap}
\end{figure}

\subsection{Results}

Once we have determined the eigenstates we are able to compute RSQA for right and left mover with different angular momentum quantum number $j$, see Fig.~\ref{fig:analyt_overlap}. The shape of the overlap has a striking similarity to Eq.\,\eqref{rsqa} derived for the cylinder geometry. By analogy, we propose the following expression for RSQA,
\begin{equation}\label{eq:overlaps_cont_disks}
\begin{split}
\mathcal{K}(j_1,j_2)~=~&\left|\int dr \psi_{-,j_2}^\dagger(r,\phi_0)\psi_{+,j_1}(r,\phi_0)\right| \\[5pt]
\approx ~& j_0^{-2}\left|j_1^2-j_2^2\right|\ ,
\end{split}
\end{equation}
where $\psi_{+,j_1}(r,\phi_0)$ is the state at total angular momentum $j_1$ moving clockwise and $\psi_{-,j_2}(r,\phi_0)$ is the state at total angular momentum $j_2$ moving counterclockwise.

We obtain again the two diagonals $j_1=j_2$ and $j_1=-j_2$, respectively, where the RSQA vanishes. When $j_1=-j_2$, the edge states are forming Kramers pairs, they have the same energy, but opposite total angular momentum. When $j_1=j_2$, the eigenstates are orthogonal by construction.

\begin{figure}[t]
\includegraphics[width=0.45\textwidth]{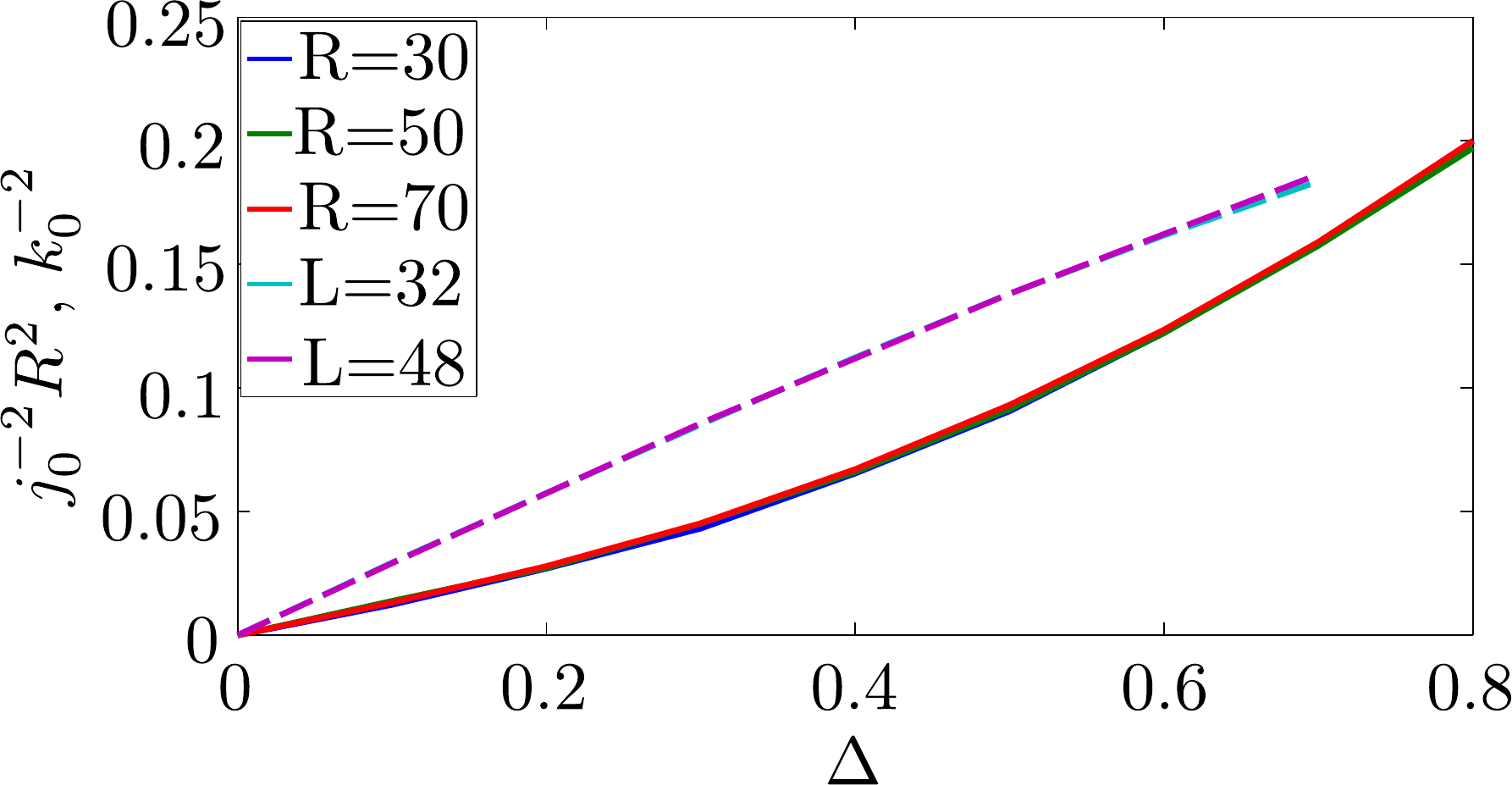}
\caption{Dependence of $j_0^{-2}R^2$ on $\Delta$ for three different disk sizes $R=30$, $50$, and $70$. In addition, the result of Fig.\,\ref{fig:k0-fit-bhz} is again shown as dashed lines to emphasize the close agreement between the two different approaches.
Parameters used: $A=5$, $B=-1$, and $M=-2$.}
\label{fig:analyt_evolution_parameter}
\end{figure}

As in the previous section, we can extract the RSQA amplitude  $j_0^{-2}$ for various values of $\Delta$. While the curves $j_0^{-2}(\Delta)$ are different for different disk sizes $R$, they agree when multiplied by $R^2$ (and are $R$ independent), see Fig.~\ref{fig:analyt_evolution_parameter}. Note that $R^2 j_0^{-2}$ has the same dimensionality as $k_0^{-2}$.  Fig.~\ref{fig:analyt_evolution_parameter} suggests that the dependence is proportional to  $|\Delta|^{3/2}$,
\begin{equation}
j_0^{-2} R^2 ~\approx~ \tilde C_{\rm BHZ} \,|\Delta|^{3/2}
\end{equation}
where $\tilde C_{\rm BHZ}$ is a constant. The $\Delta$ dependence is different compared to the results for the tight-binding nanoribbons (shown as a dashed line in Fig.~\ref{fig:analyt_evolution_parameter}). The reader may notice that there are several drastic differences which might be the source for this mismatch: (i) disk vs.\ nanoribbon and, more importantly, (ii) exact tight-binding model vs.\ continuum model. In fact, it is quite surprising that the actual values of $j_0^{-2}R^2$ are very close to the values of $k_0^{-2}$ shown in Fig.\,\ref{fig:k0-fit-bhz}.

%
%
\section{Rotation of  spin-quantization axis III: tight-binding disks}

\begin{figure}[t]
\includegraphics[width=0.45\textwidth]{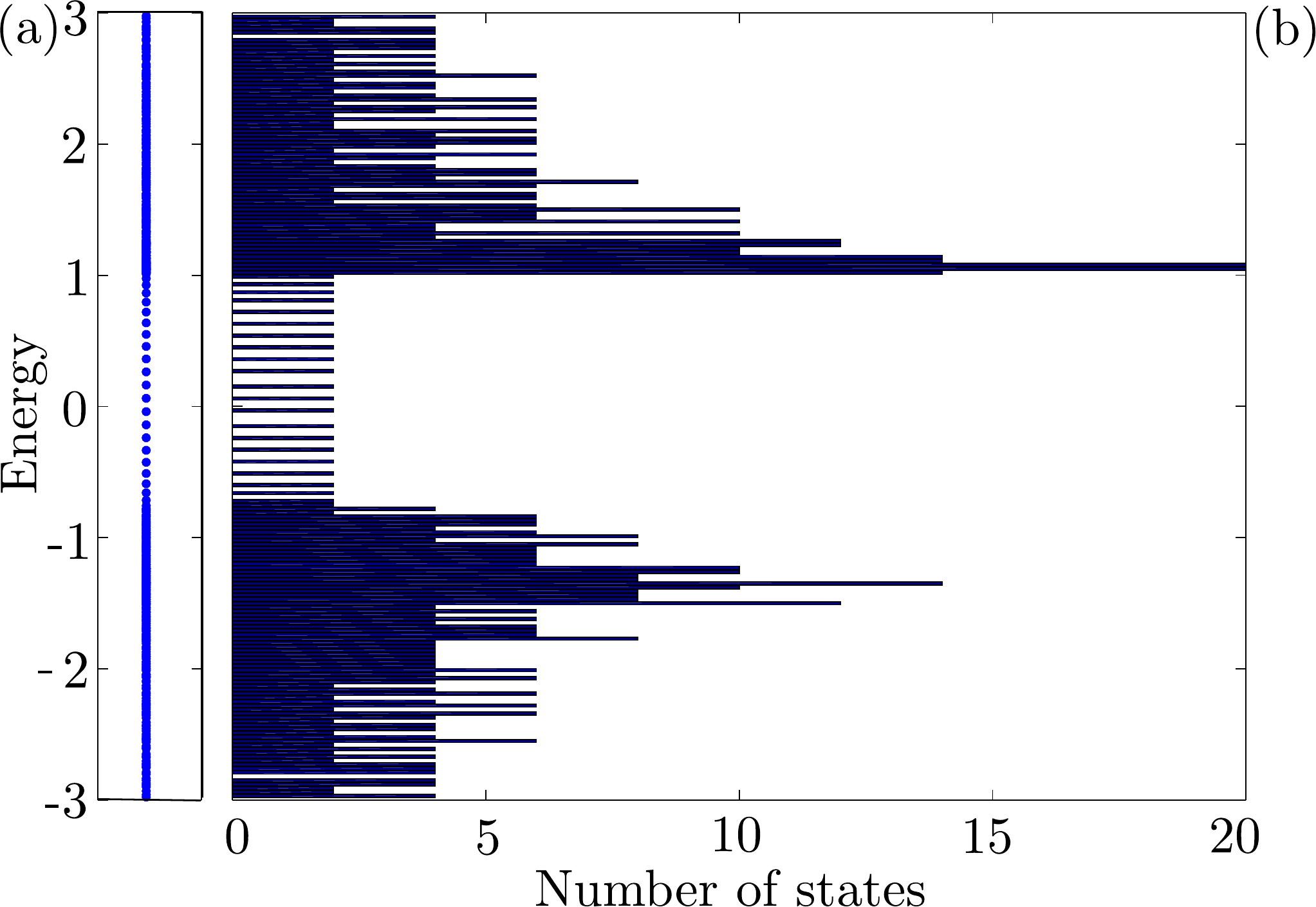}
\caption{(a) Energy levels and (b) the corresponding density of states for a KM tight-binding disk. Parameters used: $t=1$, $\lambda_{\rm SO}=0.2$, $\lambda_{\rm R,1}=0.1$, $\lambda_{\rm R,2}=0$ and $N=20\times 20=400$ sites.}
\label{fig:tightbinding_energy_dos}
\end{figure}
\begin{figure}[b]
\includegraphics[width=0.48\textwidth]{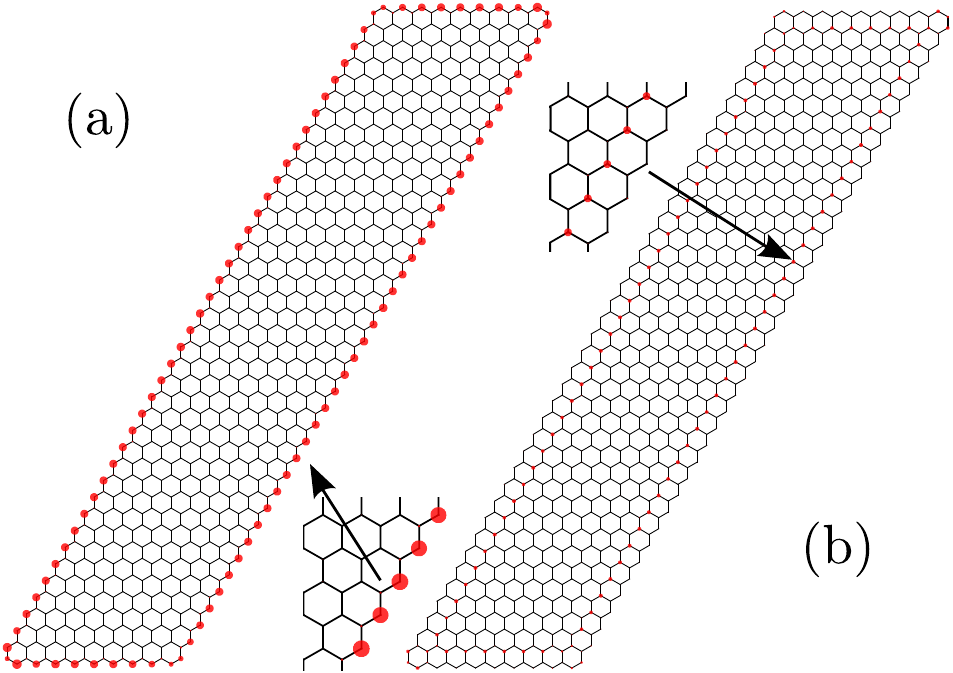}
\caption{Representative example for densities (red dots) of a single edge state for the tight binding ``disk'' .
The dot radius is proportional to the density. The  majority spin is shown in panel (a) and the minority spin in (b). Parameters used:
$t=1$, $\lambda_{\rm SO}=0.2$, $\lambda_{\rm R,1}=0.1$, and $\lambda_{\rm R,2}=0$. Here the disks possess a rectangular shape (as sketched) and $N=20\times 40=800$ lattice sites.
Note that  density for the minority spin is 20 times amplified compared to the majority spin.}
\label{fig:tightbinding_density_current}
\end{figure}
\begin{figure}[t]
\includegraphics[width=0.3\textwidth]{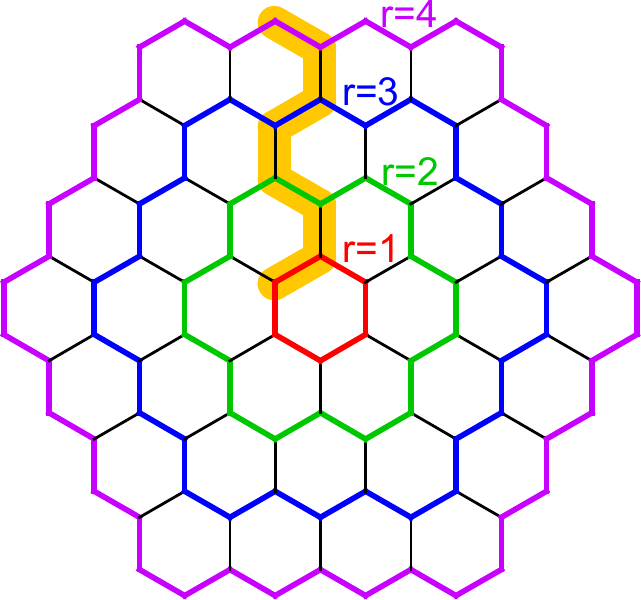}
\caption{Example for a honeycomb lattice disk with hexagonal shape. We define an effective radius $r$ as indicated by the different colors. The set $\Lambda$ of sites which is highlighted in yellow is used to compute the RSQA as discussed in the main text. The same sites are used for the spectroscopy discussion.}
\label{fig:hexagonal_lattice}
\end{figure}
\begin{figure}[t]
\includegraphics[width=0.4\textwidth]{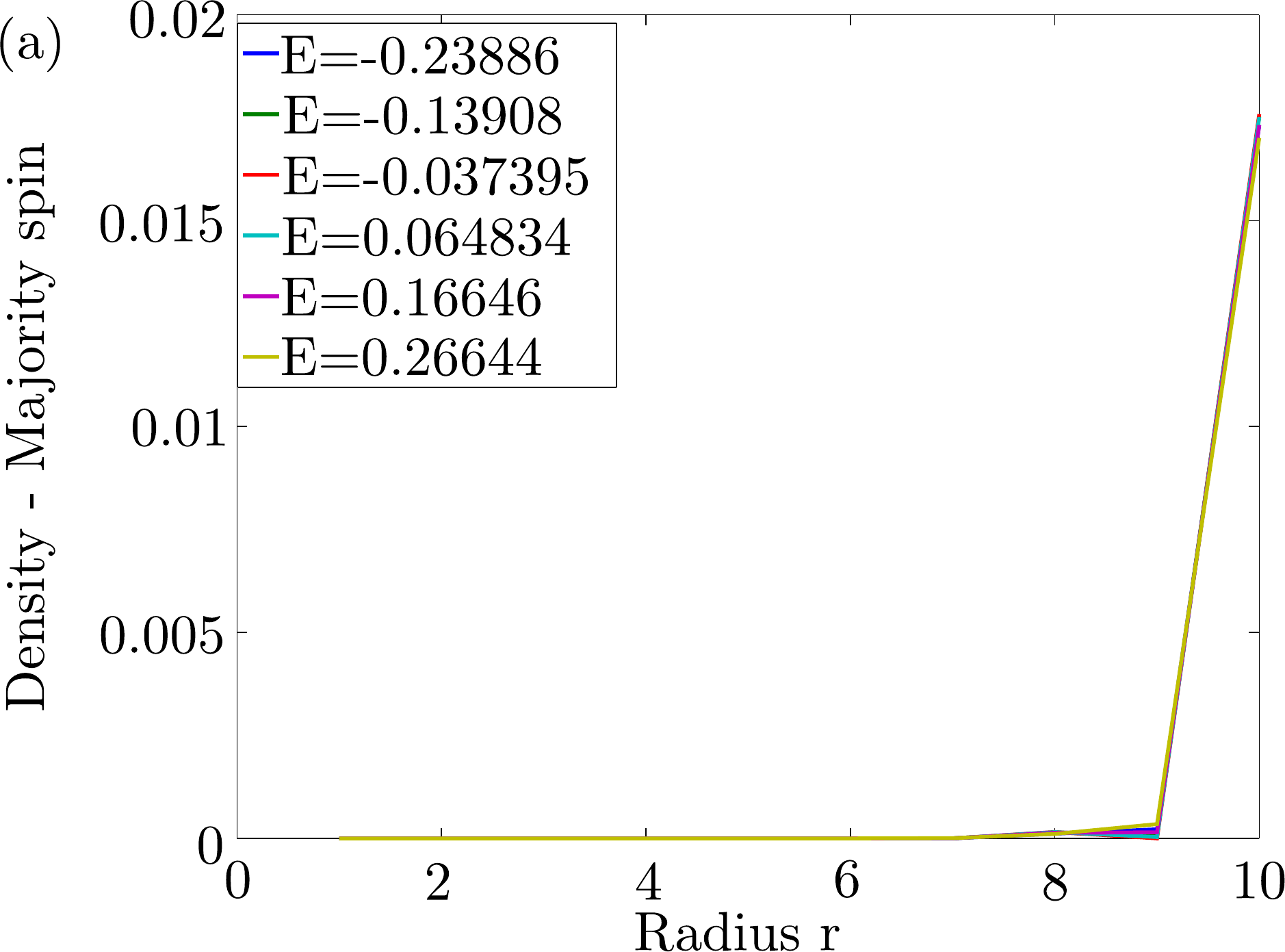}
\includegraphics[width=0.4\textwidth]{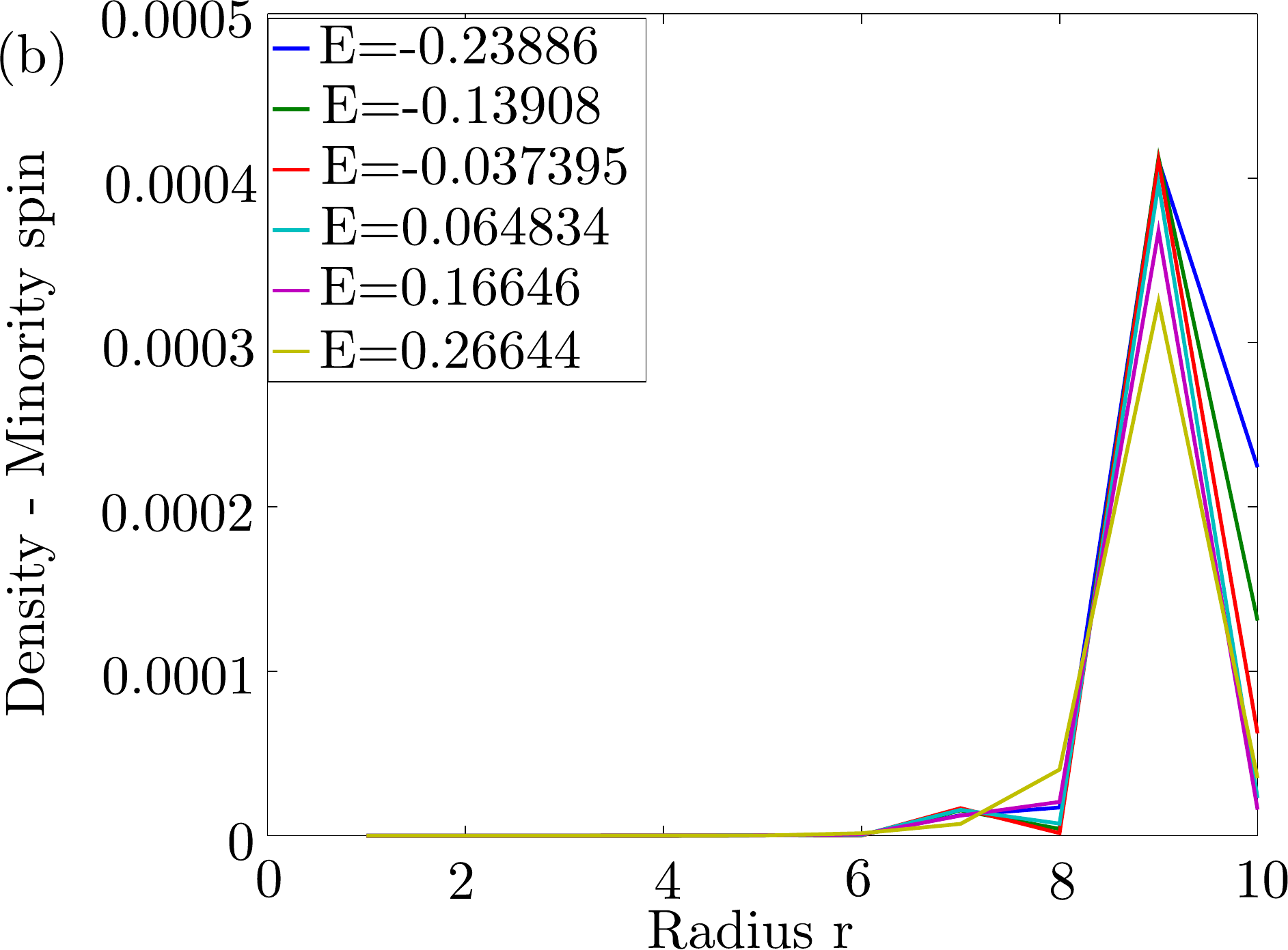}
\caption{Radial densities of a helical edges states computed on the yellow path $\Lambda$ of a tight-binding disk (see Fig.\,\ref{fig:hexagonal_lattice}). (a) Majority spin and (b) minority spin (note the different scale for the $y$ axis). Parameters used: $t=1$, $\lambda_{\rm SO}=0.2t$, $\lambda_{\rm R,1}=0.1t$, and $\lambda_{\rm R,2}=0$.}
\label{fig:tightbinding_wf}
\end{figure}
\subsection{Set-up}

In this section we consider the most general situation where neither translational nor rotational symmetry is present. We consider the tight-binding version of the KM model in real space, \ie we consider honeycomb lattice disks (or flakes). For convenience, we choose disks which have the shape of a hexagon. At the end of this section, we also consider the rectangular shaped TI samples as shown in Fig.\,\ref{fig:tightbinding_density_current} in order to demonstrate that our analysis does not rely on the shape of the samples.
Using exact diagonalization of the tight-binding Hamiltonian we can directly access energies and eigenstates. Energy spectrum and the corresponding density of states is shown in Fig.~\ref{fig:tightbinding_energy_dos}. The density of states is particle-hole asymmetric due to the presence of the first Rashba term.

Spectrum and density of states are characterized by supporting only double degenerate levels in the bulk gap which are (almost) equally spaced. We further observe the remainder of the van Hove singularity at $E/t=\pm 1$ present in the pure nearest-neighbor tight-binding model on the honeycomb lattice.

If we select one of the energy levels inside the bulk gap, we have direct access to the eigenfunction and its density distributions on the lattice. We are thus able to compute the density on the site $(i,j)$ via
\begin{equation}
\rho_\sigma(i,j)=\left|\psi_\sigma(i,j)\right|^2
\end{equation}
which is shown in Fig.~\ref{fig:tightbinding_density_current} for  a representative example.
As expected we observe a mixing of the spin channels in the presence of Rashba spin-orbit coupling.
Note that the contribution from the minority spin is much smaller than for the majority spin (for our choice of parameters by a factor 20). The same has been observed for the continuum disks discussed in the previous section.

While the wave function is mainly localized at the outermost sites (keep in mind that $\lambda_{\rm SO}=0.2$ is quite large), we observe again that the minority spin penetrates slightly deeper into the bulk (the same has been discussed for the continuum disks, see previous section).

In the following, we consider honeycomb lattice disks with a hexagonal shape as shown in Fig.\,\ref{fig:hexagonal_lattice}. While rotational symmetry is lost, this geometry comes closest to a circular shape. We can then define concentric ``rings'' around the center which have a hexagonal shape. For example, in Fig.~\ref{fig:hexagonal_lattice} the sites with the same color coding are considered to be on the same ring; they have approximately the same distance from the center of the disk.
In analogy to the continuum disks, we chose $\phi=0$ and consider only those sites which are on the yellow highlighted path in Fig.~\ref{fig:hexagonal_lattice}. A few examples for the radial densities of helical edge states are shown in Fig.~\ref{fig:tightbinding_wf}.

Due to the discreteness of the lattice the wave functions do not appear to be smooth (\eg as a function of radius), but their qualitative behavior is very similar to what was found for the continuum disks. The majority spin is strongly localized at the sample edge and its intensity dominates the wave function. As observed before, the radial density of the minority spin penetrates slightly deeper into the bulk.

\subsection{Results}

Having realized how to obtain the radial part of the wave function for the tight-binding disks we are able to compute the RSQA. In principle, we would like to apply formula \eqref{eq:overlaps_cont_disks} but for the tight-binding disks the angular momentum quantum number is not well-defined due to the lack of continuous rotational symmetry.
Furthermore, if we considered another disk shape (\eg a rectangle or a Hall bar) an approximate angular momentum number is even less justified. Nonetheless, as shown in Fig.\,\ref{fig:tightbinding_energy_dos}, each energy level inside the bulk gap is doubly degenerate, one corresponding to the right moving and the other to the left moving edge mode. Therefore we could generalize the RSQA by labeling the levels by the energy $E$ for right and left moving modes. By analogy, we therefore define the RSQA for tight-binding disks,
\begin{equation}\label{rsqa-energy}
\begin{split}
\mathcal{K}(E_1,E_2)~=~&\left|\sum\limits_{(i,j)\in\Lambda}
 \psi_{-,E_2}^\dagger(i,j)\,\psi_{+,E_1}(i,j)\right| \\[5pt]
~\approx~& \varepsilon_0^{-2}\left|E_1^2-E_2^2\right|,
\end{split}
\end{equation}
\begin{figure}[t]
\includegraphics[width=0.4\textwidth]{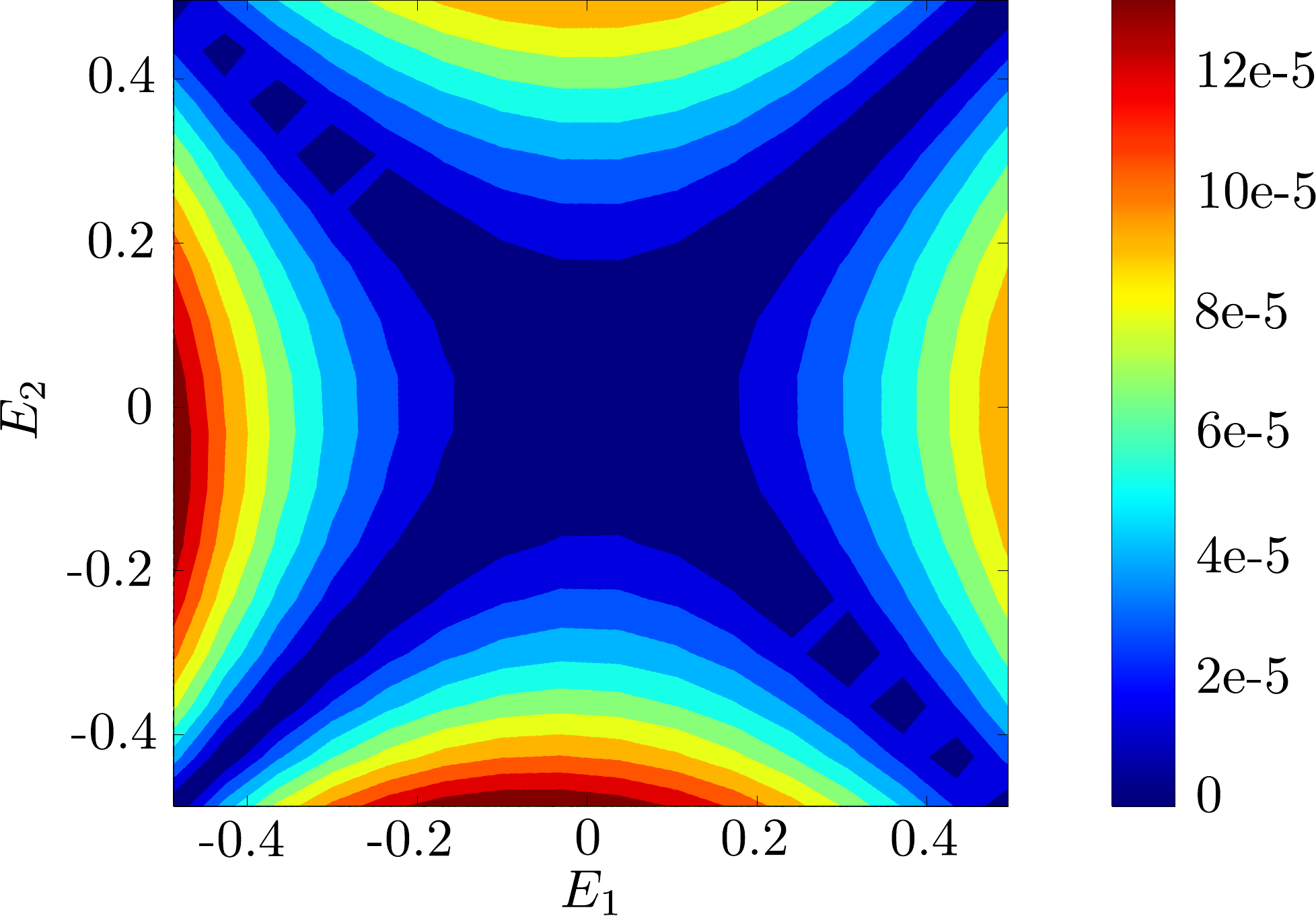}
\caption{Representative example for RSQA $\mathcal{K}(E_1,E_2)$ as a function of energie levels $E_1$ and $E_2$ of the tight-binding disk. Parameters used: $t=1$, $\lambda_{\rm SO}=0.2$ and $\lambda_{\rm R,1}=0.05$, $\lambda_{\rm R,2}=0$, and with disk size $R=15$ corresponding to 1350 lattice sites.}
\label{fig:tightbinding_overlap}
\end{figure}

\begin{figure}[t]
\includegraphics[width=0.45\textwidth]{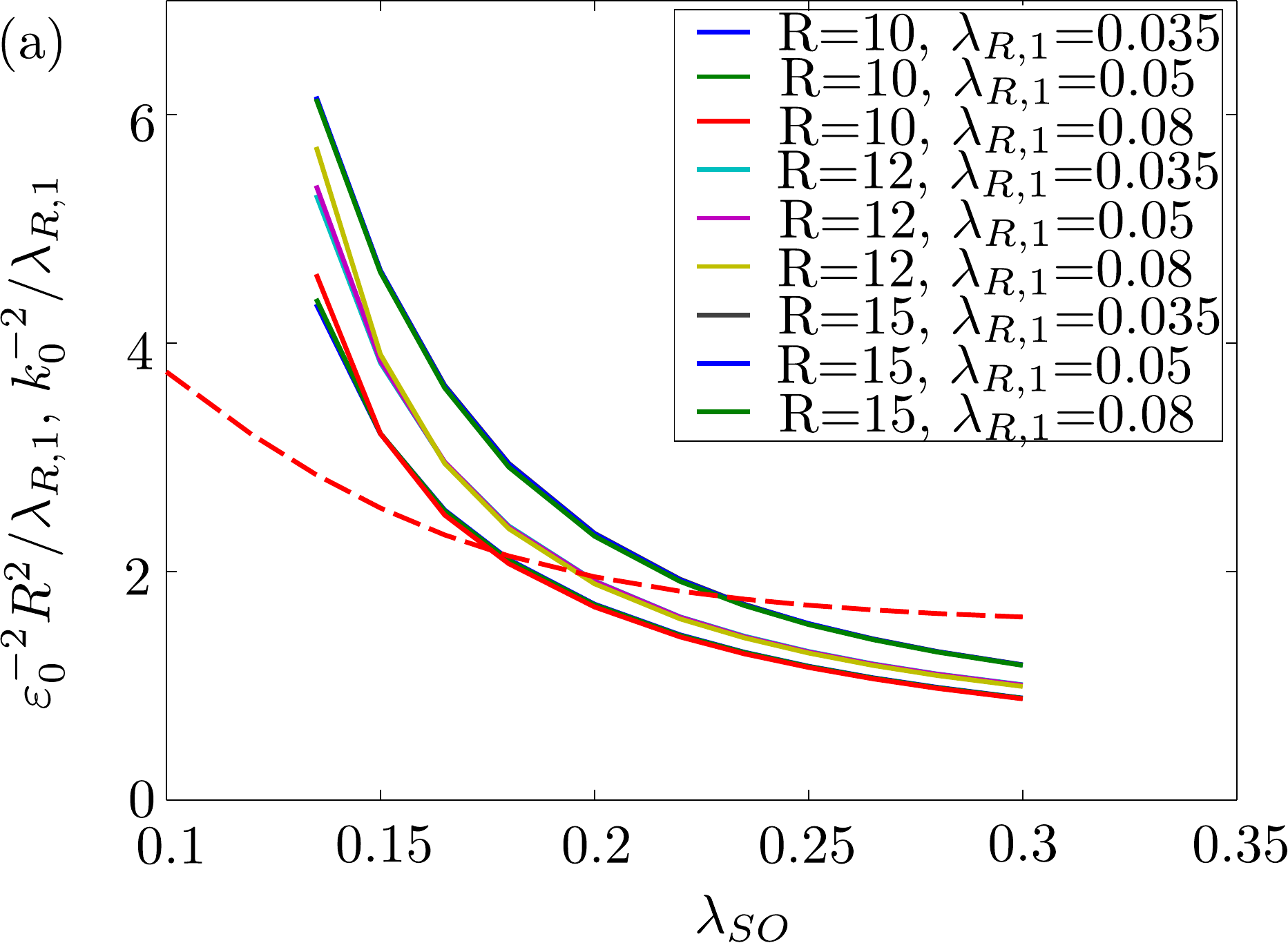}
\includegraphics[width=0.45\textwidth]{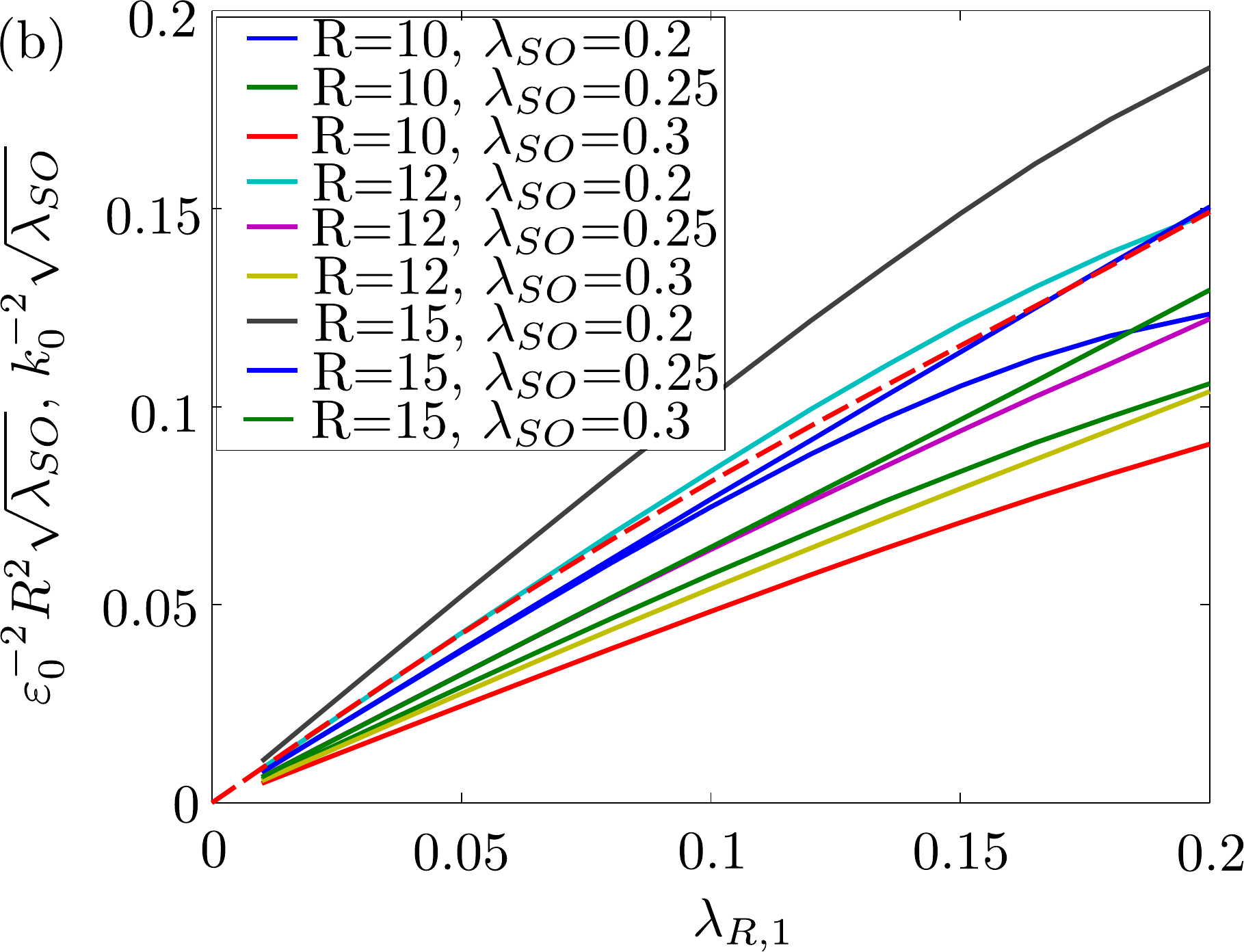}
\caption{Dependence of $\varepsilon_0^{-2}R^2$ on (a) the intrinsic spin-orbit coupling $\lambda_{\rm SO}$ and (b) the Rashba spin-orbit coupling for tight-binding disks ($t=1$) with different disk sizes $R=10$, $R=12$, and $R=15$ corresponding to $N=600$, $864$, and $1350$ lattice sites, respectively. The dashed line in panel (a) and (b) correspond to one of the curves shown in Fig.\,\ref{fig:evolution_k0} which has been rescaled by a factor 3 for comparison (see the main text for details).}
\label{fig:tightbinding_evolution_parameter}
\end{figure}
\begin{figure*}[h]
\includegraphics[width=0.3\textwidth]{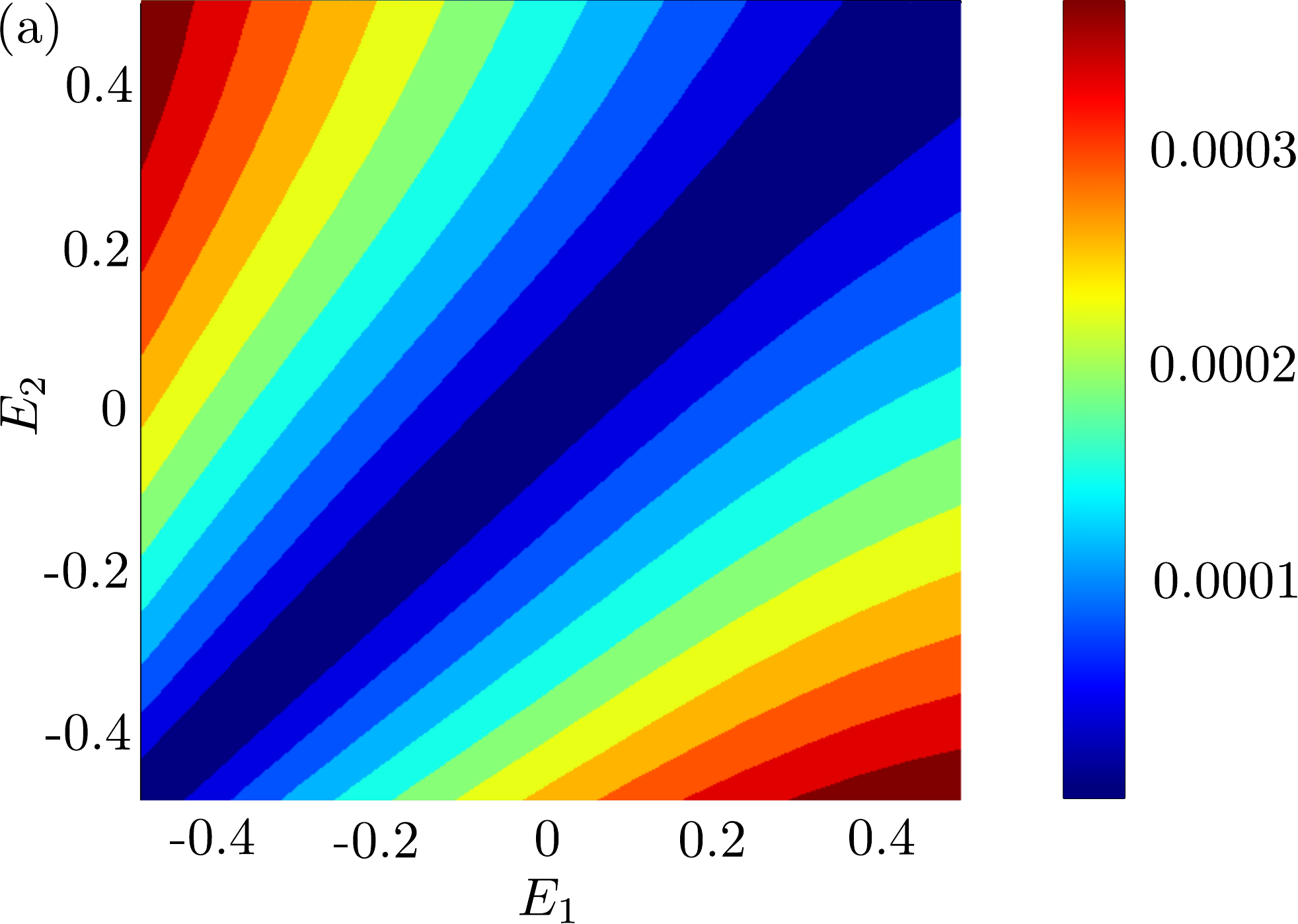}
\includegraphics[width=0.3\textwidth]{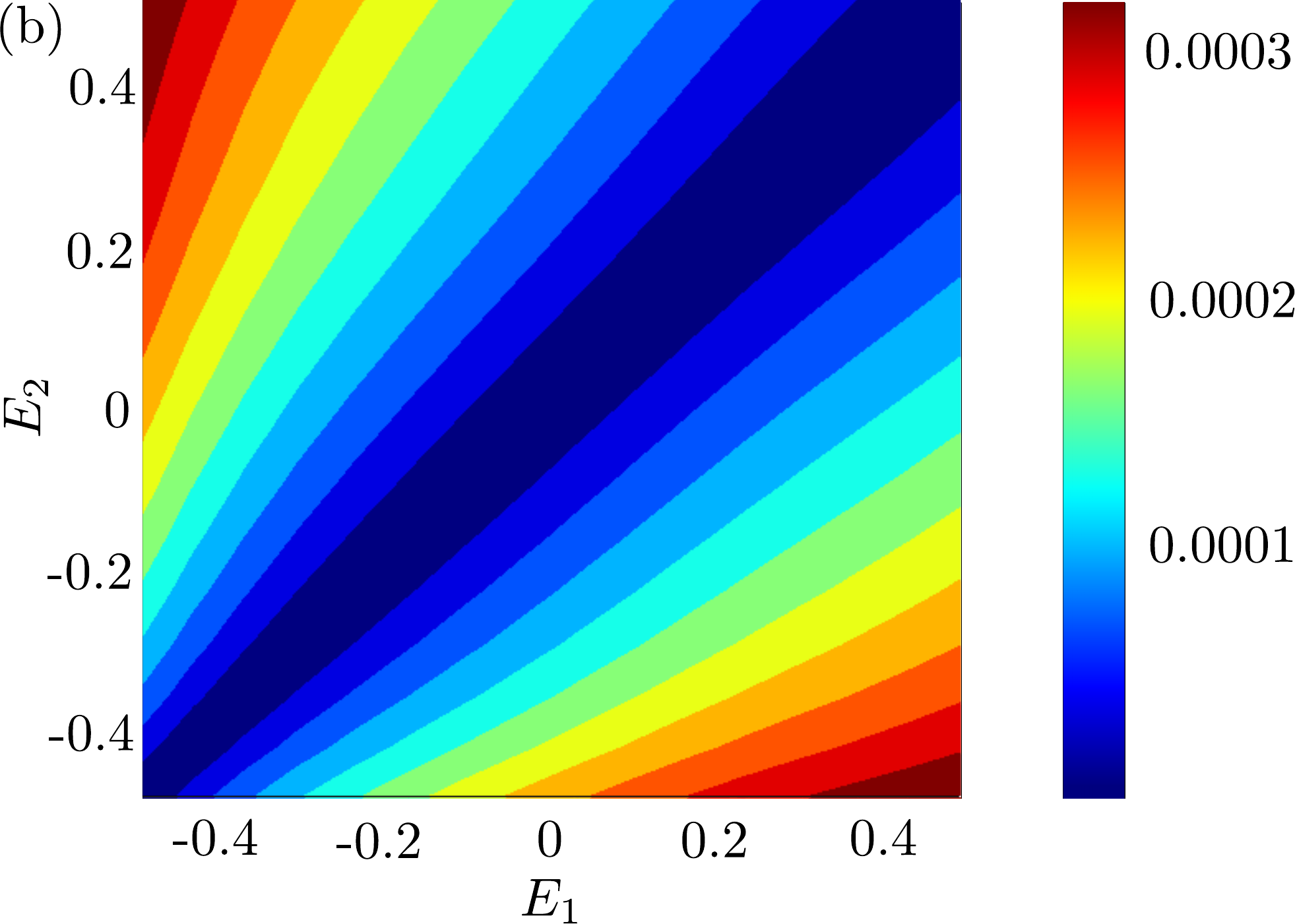}
\includegraphics[width=0.3\textwidth]{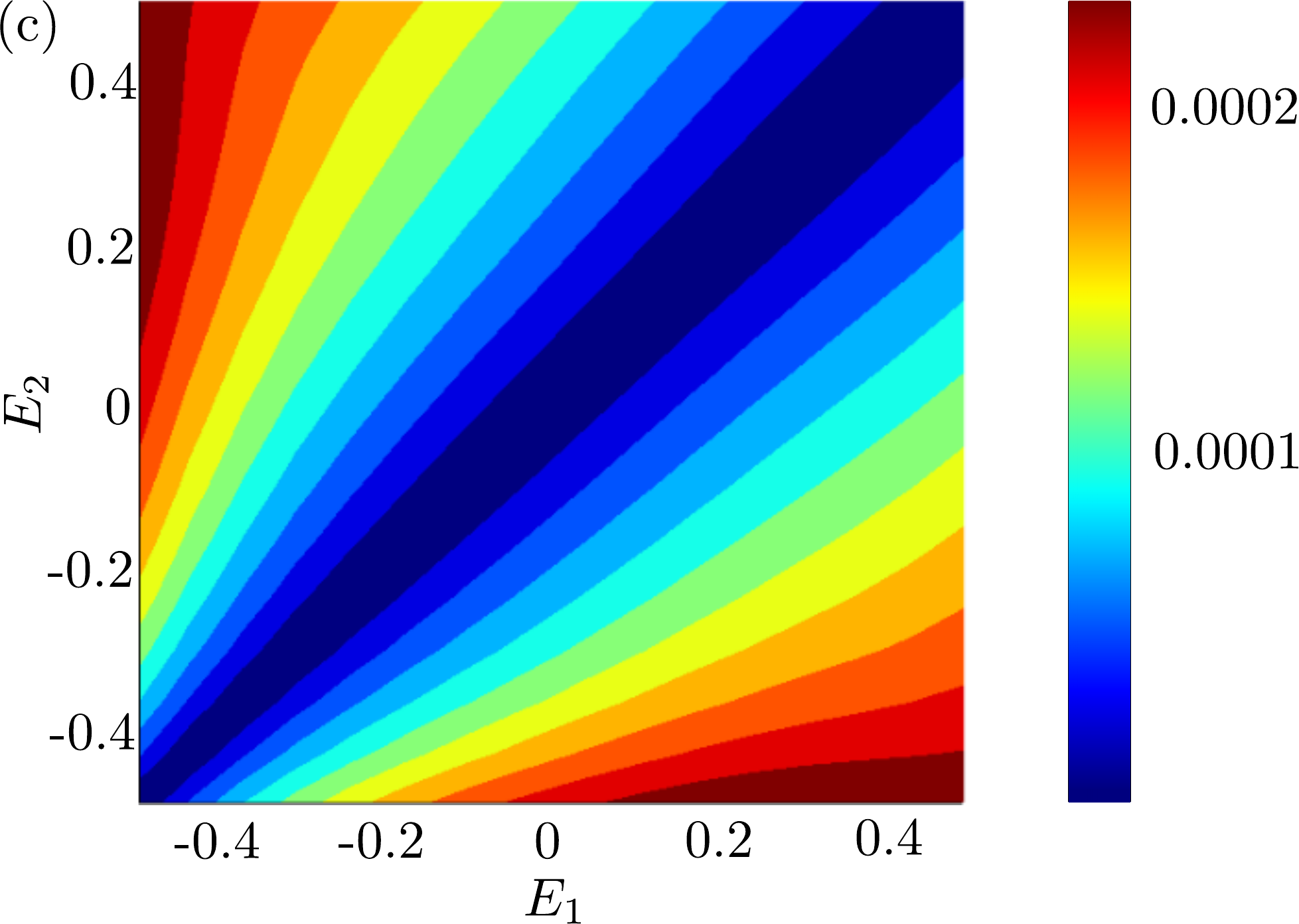}
\includegraphics[width=0.3\textwidth]{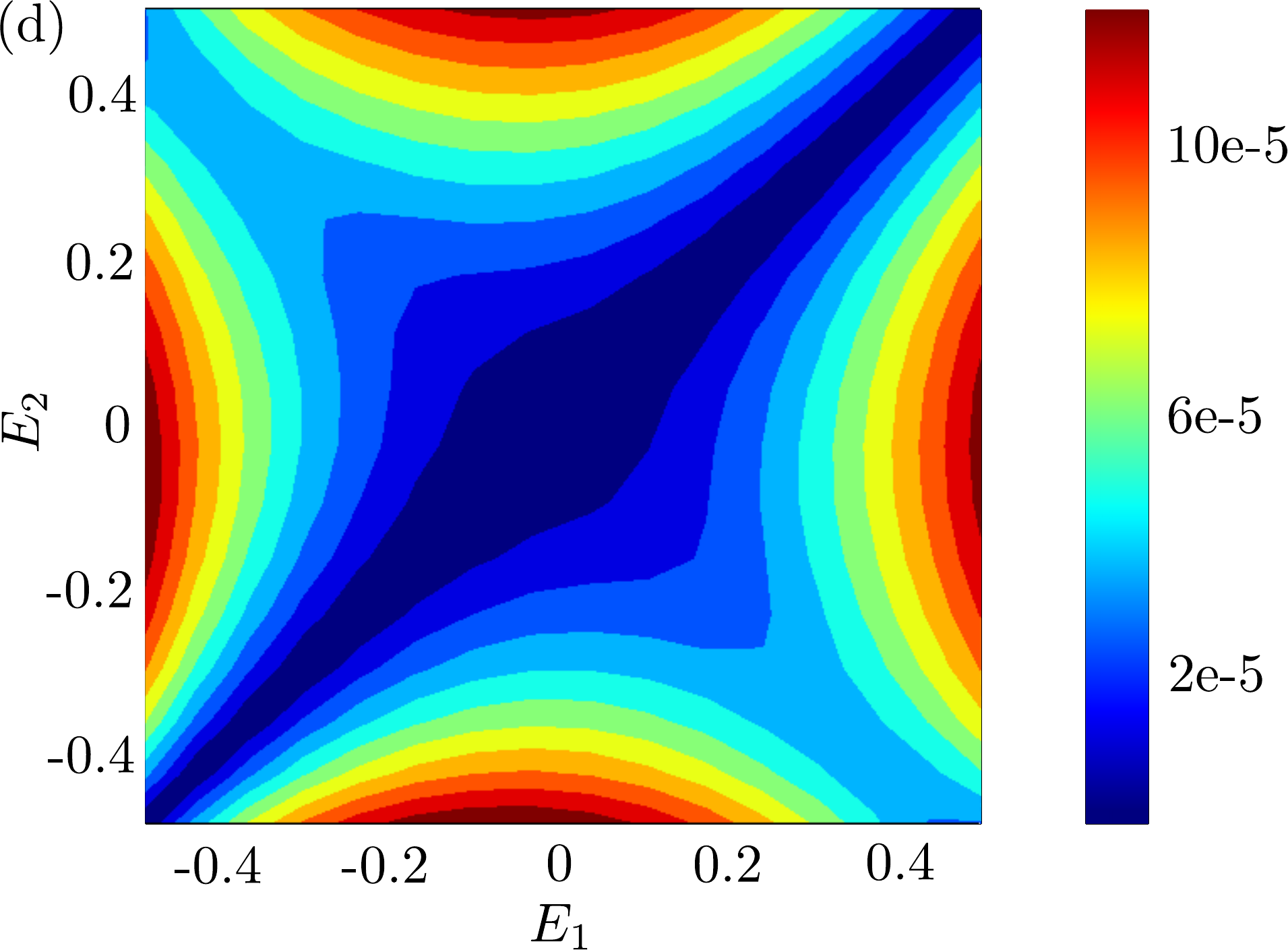}
\includegraphics[width=0.3\textwidth]{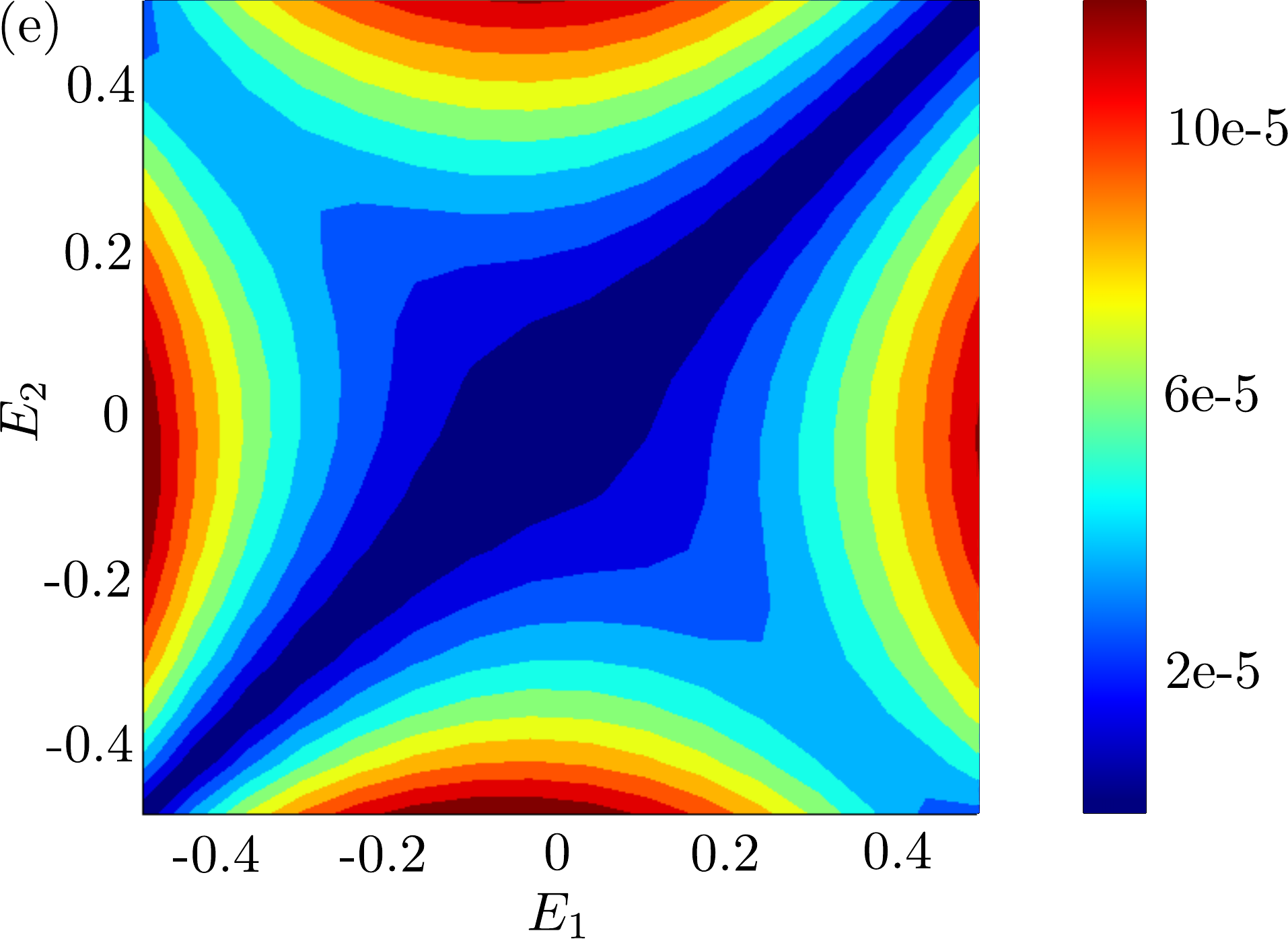}
\includegraphics[width=0.3\textwidth]{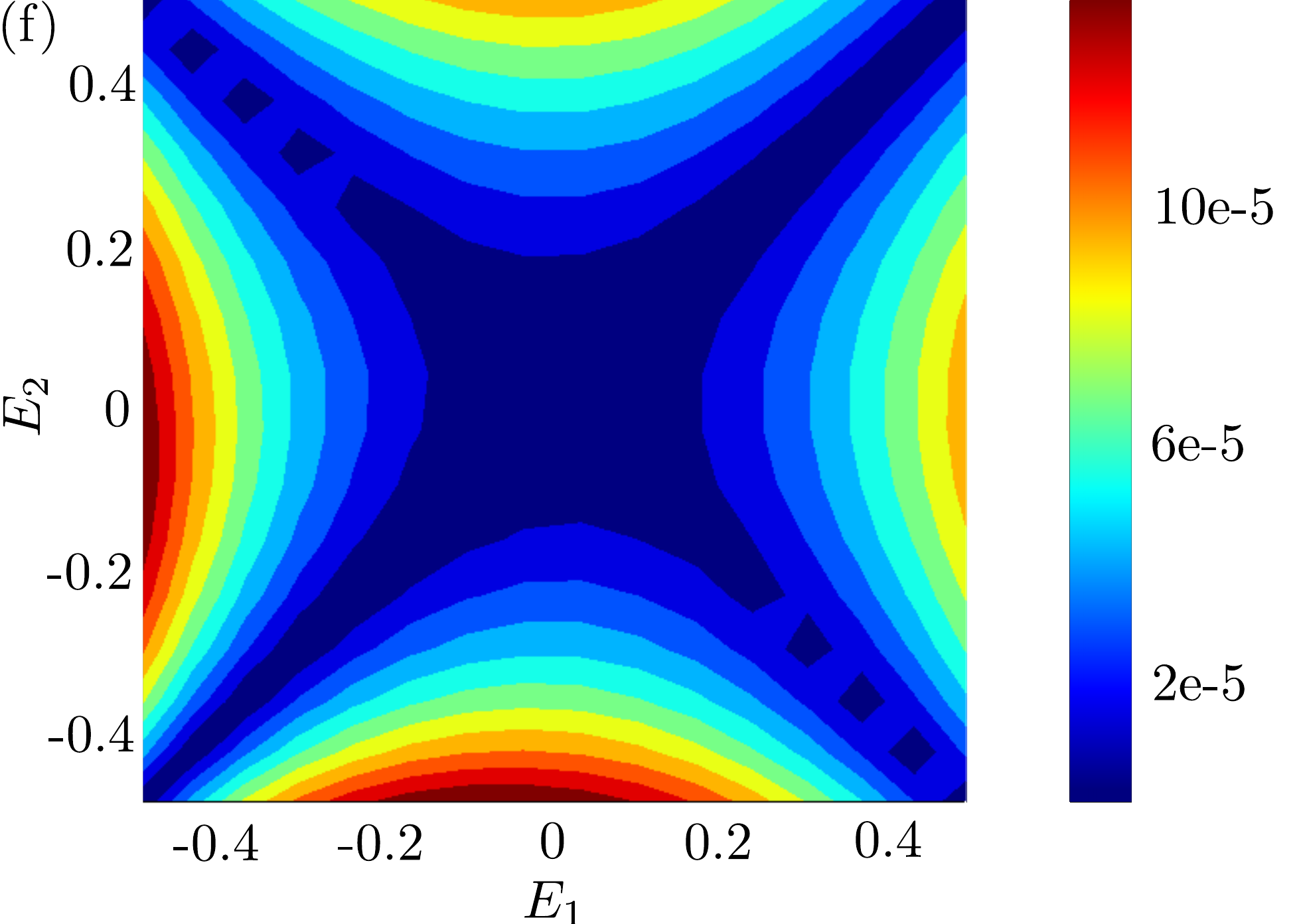}
\caption{Recovering the RSQA $\mathcal{K}(E_1,E_2)$ site-wise by ``measuring'' the wave functions only on the most outer site, the two most outer sites, etc.\ of a tight-binding disk. (a) $\mathcal{K}^{(1)}(E_1,E_2)$, (b) $\mathcal{K}^{(2)}(E_1,E_2)$, (c) $\mathcal{K}^{(3)}(E_1,E_2)$, (d) $\mathcal{K}^{(4)}(E_1,E_2)$, (e) $\mathcal{K}^{(5)}(E_1,E_2)$, and (f) $\mathcal{K}^{(6)}(E_1,E_2)$ are shown.
For the four most outer sites (d) the well-known RSQA structure becomes visible, while it is fully recovered for panel (f). Parameters used: $t=1$, $\lambda_{\rm SO}=0.2$ and $\lambda_{\rm R,1}=0.05$, $\lambda_{\rm R,2}=0$, and with disk size $R=15$ corresponding to 1350 lattice sites.}
\label{fig:spectroscopy}
\end{figure*}
where $\Lambda$ is a set of lattice sites indicated by the yellow region shown in Fig.\,\ref{fig:hexagonal_lattice} and
$\psi_{\pm,E}(i,j)$ is the wave function moving in the (counter-)clockwise direction at energy $E$, evaluated on the site $(i,j)$.
A representative example of RSQA for tight-binding disks is shown in Fig.~\ref{fig:tightbinding_overlap}.
Now we can again perform similar fitting procedures as explained in the previous sections. We extract the RSQA amplitude $\varepsilon_0^{-2}$ for different parameters and different effective radii. It turns out that the situation is comparable to the continuum disks and we should multiply the RSQA amplitude with $R^2$. In addition, we assume that the parameter dependence should be the same as in Eq.\,\eqref{k0-scaling}. Therefore we have rescaled the extracted data accordingly, see
Fig.\,\ref{fig:tightbinding_evolution_parameter}.
 We find very good agreement of the parameter dependence compared to the KM model on a nanoribbon. For illustration we have shown one of the curves from Fig.\,\ref{fig:evolution_k0} as the dashed line in Fig.\,\ref{fig:tightbinding_evolution_parameter}. Note that we have to rescale the data from the tight-binding ribbons by a factor of three because we assumed translation vectors $\bs{a}_i$ on the ribbon to have unit length such that the lenght of the Brillouin zone is $2\pi$. For the tight-binding disks we used, however, the standard length $\sqrt{3}$ between neighboring unit cells.
As RSQA depends quadratic on momentum it needs to be rescaled 
by a factor 3.

Of course, the considered tight-binding disks with hexagonal shape are a ``discrete analogue'' of a disk with circular shape and it is therefore not surprising to find such good agreement. Therefore we consider also disks with non-circular shape as shown in Fig.~\ref{fig:tightbinding_density_current}. With the parameters used in Fig.\,\ref{fig:tightbinding_density_current} we find the RSQA $\mathcal{K}(E_1,E_2)$ which looks qualitatively identical to Fig.\,\ref{fig:tightbinding_overlap}. With this final test, we have established that the RSQA will be present in any generic helical liquid independent of the considered geometry. 


\section{Spectroscopy aspect of TI disks}\label{sec:spec}
In the two previous sections we found indication of how the RSQA can be extracted in real space. In rotationally invariant systems we find the direct analogy to the nanoribbons where translational symmetry is conserved. The consideration of the tight-binding disks revealed, however, that the relevant information to compute RSQA can be locally obtained. For the tight-binding disks we showed that it is sufficient to sum up the overlap contributions from different sites along the yellow highlighted path in Fig.\,\ref{fig:hexagonal_lattice}. Since we are dealing here with edge states naively one might think that they are perfectly localized at the outermost lattice site of the disk. The numerical results show, however, that the helical edge states are typically localized on a few outer sites, although the dominant weight is indeed located at the edge site. In an idealized case it would be sufficient to measure the wave functions for different energies on a single site to obtain the RSQA directly. Using spectroscopic energy-resolved measurements that might be even accessible in experiments. Therefore we investigate now the penetration depth of the edge states in the context of RSQA. We want to understand how many sites away from the edge have to be measured to obtain RSQA. Practically, we simply modify  Eq.\,\eqref{rsqa-energy} in the following way:
\begin{equation}\label{rsqa-spectroscopy}
\mathcal{K}^{(n)}(E_1,E_2)~=~\left|\sum\limits_{(i,j)\in\Lambda_n}
 \psi_{-,E_2}^\dagger(i,j)\,\psi_{+,E_1}(i,j)\right|\,.
 \end{equation}
The only difference is that the yellow highlighted path $\Lambda$ in Fig.\,\ref{fig:hexagonal_lattice} is replaced by a subset $\Lambda_n$ of $\Lambda$. $\Lambda_1$ only contains the most outer site of $\Lambda$, $\Lambda_2$ the two most outer sites and so on. The results are shown in Fig.\,\ref{fig:spectroscopy}. We find that for the rather small disk sizes up to $R=15$ (\ie 1350 lattice sites) we need to go at least to $n=4$ in order to get qualitative agreement with the exact result. Very good agreement is obtained for $n=6$, $\mathcal{K}^{(6)}(E_1,E_2)=\mathcal{K}(E_1,E_2)$. Note that a disk with size $R=15$ is still a very small sample. Typical samples sizes in experiments are magnitudes larger. This means that in real experiments it might be sufficient to measure the outermost two or three sites to obtain perfect agreement with the full RSQA. This opens the perspective towards a direct spectroscopic measurement of RSQA in a realistic experiments. Details of this spectroscopy aspect will be worked out in the future.

\section{Conclusion}\label{sec:conclusions}
We have analyzed the spin texture of generic helical liquids, the gapless helical edge states of two-dimensional topological insulators with broken axial spin symmetry. Generic helical liquids feature a momentum-dependent rotation of the spin-quantization axis which eventually is responsible for the leading finite-temperature correction to the otherwise quantized Hall conductance.\cite{schmidt-12prl156402}
Here we considered two different topological insulator models, the Bernevig--Hughes--Zhang and the Kane--Mele model, and computed the rotation of the spin-quantization axis for these models in detail. For systems with translation invariance, we considered parameter and sample size dependencies. Moreover, we considered disks which do not possess translational symmetry anymore but rotational symmetry instead. We showed that in such disks the spin-quantization axis rotates as a function of total angular momentum. Finally, we showed that the rotation of the spin-quantization axis remains accessible even when neither continuous translational nor rotational symmetries are preserved. This proves the ubiquity of the rotation of spin-quantization axis, which is independent of the considered geometry. The study of tight-binding disks or flakes also revealed that the information needed can be extracted almost locally, opening the path towards spectroscopic detection of the spin texture of helical edge states. 


\begin{acknowledgments}
This work was supported through the DFG priority program SPP 1666 ``Topological insulators''. AR 
acknowledges support by the DFG through SFB 1143.
TLS acknowledges support by the Swiss National Science Foundation and the National Research Fund, Luxembourg (ATTRACT 7556175).
SR is supported by the DFG through FOR 960, through SFB 1143, and by the Helmholtz association through VI-521.
\end{acknowledgments}

\bibliographystyle{prsty}
\bibliography{paper}

\end{document}